\documentclass[11pt,a4paper]{article}
\usepackage{jcappub}
\usepackage{float}
\usepackage{rotating}
\usepackage{graphicx}
\usepackage{epsfig}
\usepackage{mathtools}
\usepackage{amsmath}
\usepackage{amssymb}
\usepackage{mathrsfs}  
\usepackage{subfigure}
\usepackage{time}
\usepackage{orcidlink}
\usepackage{relsize, comment}
\usepackage{xcolor}
\usepackage[utf8]{inputenc}

\usepackage{booktabs}

\hypersetup{ linktoc=all,
    colorlinks=true, linkcolor={green},  
       citecolor={red}, urlcolor={vividviolet}
}
\definecolor{rosy}{RGB}{230,235,252}
\definecolor{myframetitle}{RGB}{90,89,170}
\definecolor{myblocktitle}{RGB}{140,185,249}
\definecolor{mytitle}{RGB}{10,80,26}

\definecolor{darkgreen}{RGB}{27,130,45}
\definecolor{darkblue}{rgb}{0,0,0.3}
\definecolor{darkred}{rgb}{0.7,0,0}

\definecolor{light gray}{RGB}{220,220,220}
\definecolor{dark purple}{RGB}{108,0,217}
\definecolor{pink}{RGB}{190,20,100}
\definecolor{orang}{RGB}{193,63,0}
\definecolor{green}{RGB}{11,98,17}
\definecolor{darkpink}{RGB}{153,0,76}
\definecolor{bluegreen}{RGB}{0,102,102}
\definecolor{greenlagan}{RGB}{0,102,0}
\definecolor{redgreen}{RGB}{102,102,0}
\definecolor{Redgreen}{RGB}{153,76,0}
\definecolor{vividviolet}{rgb}{0.62, 0.0, 1.0}
\definecolor{amaranth}{rgb}{0.9, 0.17, 0.31}
\definecolor{palatinateblue}{rgb}{0.15, 0.23, 0.89}
\definecolor{brightpink}{rgb}{1.0, 0.0, 0.5}
\definecolor{cornflowerblue}{rgb}{0.39, 0.58, 0.93}
\definecolor{deepcarminepink}{rgb}{0.94, 0.19, 0.22}
\definecolor{radicalred}{rgb}{1.0, 0.21, 0.37}

\usepackage{bm}

\title{Cubic Curvature Deformations to the Kerr Black Hole Shadow
}
\author[a]{Alireza Allahyari\orcidlink{0000-0002-4553-2436}}
\emailAdd{alireza.al@khu.ac.ir}
\affiliation[a]{Department of Astronomy and High Energy Physics, Kharazmi University, 15719-14911, Tehran, Iran \looseness=-1}

\author[b,c,d]{Shinji Mukohyama}%
\emailAdd{shinji.mukohyama@resceu.s.u-tokyo.ac.jp}
\affiliation[b]{Research Center for the Early Universe (RESCEU), Graduate School of Science, The
University of Tokyo, Hongo 7-3-1, Bunkyo-ku, Tokyo 113-0033, Japan
}
\affiliation[c]{Center for Gravitational Physics and Quantum Information,
Yukawa Institute for Theoretical Physics, Kyoto University, 606-8502, Kyoto, Japan}
\affiliation[d]{Kavli Institute for the Physics and Mathematics of the Universe (WPI),
The University of Tokyo Institutes for Advanced Study,
The University of Tokyo, Kashiwa, Chiba 277-8583, Japan}

\author[e,f]{Yi Pang}
\emailAdd{pangyi1@tju.edu.cn}
\affiliation[e]{Center for Joint Quantum Studies and Department of Physics,
School of Science, Tianjin University, 135 Yaguan Road, Tianjin 300350, China}
\affiliation[f]{Peng Huanwu Center for Fundamental Theory, Hefei, Anhui 230026, China}

\abstract{
We investigate the effects of cubic curvature corrections on the shadows of rotating black holes within the effective field theory of gravity. Starting from a perturbatively corrected Kerr metric incorporating two independent cubic curvature couplings, we construct black hole shadows through numerical integration of null geodesics, without assuming separability of the geodesic equations. Using a Hamiltonian ray-tracing approach, we systematically examine the dependence of the shadow geometry on the higher-curvature couplings, black hole spin, and observer inclination. We find that cubic curvature interactions can produce discernible deviations from the Kerr shadow, modifying its apparent size and shape. In particular, the two cubic interactions exhibit distinct effects on the shadow boundary, with the magnitude of the deviations depending sensitively on the coupling strengths and black hole spin. We validate our numerical results through independent convergence tests of the geodesic integration and shadow-boundary determination, and recover the analytical Kerr shadow in the vanishing-coupling limit. Our results provide a complementary probe of higher-curvature gravity in the strong-field regime and illustrate the potential of black hole shadow observations to constrain departures from general relativity.
}

\begin{document}
\begin{flushright} {\footnotesize RESCEU-31/26, YITP-26-162, IPMU26-0039, USTC-ICTS/PCFT-26-82} \end{flushright}
\maketitle
\flushbottom
\section{Introduction}
Black holes are among the most remarkable predictions of general relativity. Oppenheimer and Snyder first demonstrated that gravitational collapse can lead to the formation of black holes \cite{oppenheimer}, while the singularity theorems established that trapped surfaces and spacetime singularities arise under generic physical conditions \cite{penrose65,hawkingpenrose}. Subsequent uniqueness theorems showed that stationary  black holes are fully characterized by a small set of parameters: mass, angular momentum, and electric charge \cite{israel,carter71,hawking72,robinson,carter68}. These theoretical developments laid the foundation for the modern understanding of black holes, as presented in standard references \cite{chandra,mtw,wald,frolovnovikov}. Today, black holes are firmly established as astrophysical objects, supported by observations across multiple scales: stellar-orbit measurements constrain the mass of the compact object at the Galactic Center \cite{ghez08,genzel10,gravity18}, gravitational-wave detections have revealed mergers of stellar-mass black holes \cite{gw150914}, and dynamical studies of gas and stars have provided mass estimates for the supermassive black hole at the center of M87 \cite{gebhardt11,walsh13}.

A black hole is not directly visible, but it leaves a characteristic imprint on light from its surroundings. Strong lensing deflects photons that pass close to the hole \cite{einstein36,darwin,ohanian}, and photons with impact parameters below a critical value are captured by the horizon. For a distant observer, the capture region appears as a dark region against a bright background, the \emph{black hole shadow}. The first treatments of photon capture were given by Synge for the Schwarzschild case \cite{synge} and by Bardeen for Kerr, where the shadow boundary is a deformed curve set by the spin and inclination \cite{bardeen73,bpt72}. Early images of accreting black holes \cite{cunninghambardeen,luminet} and the proposal to image the shadow of the Galactic Center black hole \cite{falcke00} made the effect an observational target. The shadow edge is determined by the bound photon orbits of the spacetime \cite{teo,devries,cunha17}, and the size of a Schwarzschild shadow is $3\sqrt{3}\,GM/c^{2}$ in radius. The theory has been developed extensively, including the Kerr and charged cases and their dependence on spin and inclination \cite{hiokimaeda,grenzebach,tsukamoto18,tsupko17,bambifreese,takahashi,zakharov05a,zakharov05b}, and is summarized in the reviews \cite{perlicktsupko22,cunhaherdeiro18,dokuchaev}. The relation between the shadow, the photon ring and the observed image is clarified in \cite{gralla19,bronzwaerfalcke,narayan19,gralla20a,gralla20b,beckwithdone}. The strong-field lensing that structures this image has been analyzed in \cite{virbhadraellis00,bozza01,bozza02,eiroa02,claudel,perlick04,bozza10}.

The observation of the shadow became possible with very long baseline interferometry at millimeter wavelengths, where the emission near the horizon is optically thin. Early detections of horizon-scale structure \cite{doeleman08,doeleman12,akiyama15} led to the Event Horizon Telescope (EHT) images of M87$^{*}$ \cite{eht1,eht2,eht3,eht4,eht5,eht6} and of Sagittarius~A$^{*}$ \cite{sgra1,sgra2,sgra3,sgra4,sgra5,sgra6}. Subsequent work reported the polarized structure of M87$^{*}$ \cite{eht7,eht8}, with theoretical studies of polarized images from magnetized Kerr black holes \cite{lee23},
 the persistence of its ring across epochs \cite{eht2018}, a connection to the jet base \cite{lu23}, and the variability of Sgr~A$^{*}$ \cite{wielgus22}. These results rely on advanced imaging methods \cite{akiyama17,chael18} and are interpreted using general relativistic magnetohydrodynamic simulations \cite{harm,porth19,dexter10,moscibrodzka16} combined with ray tracing and radiative transfer \cite{gyoto,gold20}. The underlying accretion and jet physics is described in \cite{shakura,narayanyi,yuannarayan,abramowicz,bz77,blandford19}, and earlier studies of near-horizon hot spots anticipated features of the observations \cite{broderickloeb05,broderickloeb06}. Future arrays, including space baselines, aim to resolve the sharp photon ring predicted by theory \cite{johnson20,hadar21,johnson23,doeleman23,roelofs19}.

Because the shadow depends on the geometry of spacetime, it provides a new tool for testing gravity in the strong-field regime. For Kerr black holes the shadow is nearly circular and its size is fixed by the mass and distance, which allows tests of the no-hair property \cite{johannsenpsaltis10,johannsen13,johannsen16b,broderick14,johannsen16a,psaltis19,bambi17}. The EHT measurements were used to constrain deviations from the Kerr metric \cite{psaltis20,sgra6,bambi19,mizuno18,vagnozzi23,ehtcharge}. These studies use parametrized metrics \cite{rezzolla14,konoplya16,younsi16,atamurotov13} and specific alternatives: scalar hair and other hairy solutions \cite{cunha15,cunha19,khodadi20,Meng:2024puu, Qin:2022kaf,Chen:2024cxi}, nonlinear electrodynamics and regular black holes \cite{allahyari20,abdujabbarov16,amarilla12}, modified gravity and quantum-inspired geometries \cite{amarilla10,kumarghosh20b,held19,Peng}, and dark matter environments \cite{jusufi19,konoplya19}. Horizonless and exotic compact objects, naked singularities and wormholes give different shadows and lensing signatures \cite{virbhadraellis02,shaikh18,shaikhjoshi,nedkova,ohgami,abdikamalov,cardosopani,cardoso16}. Further effects include plasma dispersion \cite{perlick15,perlick17}, a cosmological constant \cite{perlick18}, and use of the shadow as a standard ruler \cite{tsupko20}. Methods to infer parameters from the shadow shape are given in \cite{kumarghosh20a,wei19}. A caveat, emphasized in the literature, is that astrophysical uncertainties in the emission region limit how cleanly a shadow can be mapped to a metric \cite{bronzwaerfalcke,narayan19,sgra5}.

Testing general relativity in the strong-field regime is motivated by the expectation that gravity may be modified at high curvature \cite{berti15}, and the no-hair property of black holes is a central target of such tests \cite{cardosogualtieri16}. A systematic way to parametrize these modifications is the effective field theory (EFT) of gravity \cite{burgess04,endlich17}, whose higher-curvature operators change black hole perturbations and quasinormal modes \cite{derham20,cano20,cano22}, and, in the cubic case, the black hole solutions themselves \cite{bueno16,hennigar17}. Accreting black holes offer a complementary probe: quasi-periodic oscillations (QPOs) can be interpreted through the frequencies of orbital motion in the strong field \cite{stella99}, which in GRO J1655$-$40 yields precise mass and spin measurements \cite{motta14}. For shadows, deviations from Kerr are partly degenerate with the spin \cite{li14,tsukamoto14,ghasemi15,wei13}, and separating gravitational from astrophysical effects requires care about the emission model \cite{gralla21,kocherlakota22,chael18heating}. Photon-ring features on long baselines, including their polarization, are less sensitive to this uncertainty and may provide cleaner tests \cite{gralla20c,himwich20}, while the use of shadow sizes as cosmological rulers faces its own concerns \cite{vagnozzi20ruler}.

In the framework of effective field theory, higher-curvature interactions introduce corrections to the Einstein–Hilbert action that can modify the geometry of black holes and, consequently, the propagation of photons in their vicinity. In four-dimensional spacetime, the leading nontrivial corrections to the Kerr solution arise from cubic curvature terms, which can alter the locations of unstable circular photon orbits and the resulting shadow boundary.  Cubic curvature interactions can arise from two-loop quantum corrections to Einstein gravity \cite{Goroff:1985th} or from non-supersymmetric string theory \cite{Metsaev:1986yb}. Since such interactions are forbidden even in theories with minimal
supersymmetry in four dimensions \cite{Ozkan:2024euj}, investigating their physical consequences may provide distinctive signatures of quantum gravity theories without underlying supersymmetry.

 In our previous work, we obtained the cubic curvature corrected Kerr metric up to first order in the cubic couplings and perturbative in the rotation parameter analytically \footnote{The corrected Kerr solution non-perturbative in rotation parameter was recently obtained in \cite{Fernandes:2025vxg} numerically}, and investigated their observational implications using quasi-periodic oscillations from accreting black hole systems~\cite{Allahyari:2025bvf}.  Motivated by these developments, studying the shadow of this higher-curvature-corrected black hole can provide complementary insight into the effects of strong gravitational fields on photon trajectories and the possible observational signatures of departures from the Kerr spacetime. In particular, analyzing the shadow's apparent size, shape, and deformation as functions of the black hole spin and higher-curvature coupling parameters may help characterize the impact of cubic curvature corrections and assess the potential of black hole imaging as a probe of effective theories of gravity.

For a general stationary and axisymmetric metric, a direct ray-tracing construction has an important advantage: it does not require an a priori assumption that the null geodesic equations are separable. The observable is obtained from the spacetime metric itself by launching rays from the observer, integrating them through the geometry, and determining whether they are captured or escape. This approach is particularly appropriate for the Kerr-cubic geometry considered here, whose metric coefficients contain nontrivial radial and angular deformation functions.

The purpose of this work is twofold. First, we give a self-contained description of the numerical construction used to obtain the shadow. Second, we establish a numerical error budget that is directly connected to the observable. The latter point is important because a visually smooth shadow is not, by itself, evidence of convergence. Two numerical operations enter at different stages of the calculation: the integration of each null geodesic and the localization of the capture/escape transition on each radial screen ray. We therefore test these two sources of error independently.

We use dimensionless variables throughout. The dimensionless radial coordinate is denoted by $\rho=r/M$, the dimensionless spin by $\chi=a/M$, and the two dimensionless deformation parameters are introduced in section~\ref{sec:metric}. The numerical implementation uses the exact polynomial-rational expressions supplied for the five deformation functions. Those expressions are lengthy; rather than reproducing them in a form that could introduce transcription errors, we regard the accompanying machine-readable implementation as the defining representation of the metric functions.

The paper is organized as follows. Section~\ref{sec:metric} specifies the metric and conventions. Section~\ref{sec:shadow} describes the finite-distance ray-tracing calculation, the local observer frame, the geodesic equations, and the construction of the shadow boundary. Section~\ref{sec:numerics} gives the numerical error analysis, including the independent DOP853 and boundary-root convergence tests. Sections~\ref{sec:b5}--\ref{sec:inclination} describe the parameter scans in the two cubic deformations and in observer inclination. We conclude in section~\ref{sec:discussion}.

\section{Metric with Cubic Curvature Deformations}\label{sec:metric}
We work in Boyer--Lindquist-like coordinates $(t,r,\theta,\phi)$. The mass scale is denoted by $M$, and the dimensionless radial coordinate and spin parameter are
\begin{equation}
 \rho=\frac{r}{M},\qquad
 \chi=\frac{a}{M}.
\end{equation}
In the numerical calculations the mass scale is set to unity, so that $\rho$ is represented by the variable $r$ in the code. 
It is convenient to introduce
\begin{equation}
 x=\cos\theta,
 \qquad
 \Sigma=r^2+\chi^2x^2,
 \qquad
 \Delta=r^2-2r+\chi^2,
\end{equation}
where the final expressions are written in the dimensionless variables used by the numerical implementation. The deformation functions are denoted by $H_i(r,x,\chi, \beta_5,\beta_6)$, with $i=1,\ldots,5$~ \cite{Allahyari:2025bvf}, where $\beta_5$ and $\beta_6$ are the coefficients of $RR_{\mu\nu\rho\lambda}R^{\mu\nu\rho\lambda}$ and $R^{\mu\nu}{}_{\lambda\rho}R^{\lambda\rho}{}_{\sigma\delta}R^{\sigma\delta}{}_{\mu\nu}$ terms respectively, as other independent cubic terms will not affect the solution. The nonvanishing covariant metric components are
\begin{align}
 g_{tt}&=-\left(1-\frac{2r}{\Sigma}-H_1\right),\\
 g_{t\phi}&=-\frac{2\chi r\sin^2\theta}{\Sigma}(1+H_2),\\
 g_{rr}&=(1+H_3)\frac{\Sigma}{\Delta},\\
 g_{\theta\theta}&=(1+H_5)\Sigma,\\
 g_{\phi\phi}&=(1+H_4)
 \left(r^2+\chi^2+\frac{2\chi^2r\sin^2\theta}{\Sigma}\right)\sin^2\theta.
\end{align}
The undeformed limit $\beta_5=\beta_6=0$ recovers the Kerr metric. 
The inverse metric is evaluated without a numerical matrix inversion at every geodesic step. The $(t,\phi)$ block is inverted analytically. In particular, if
\begin{equation}
 D=g_{tt}g_{\phi\phi}-g_{t\phi}^2,
\end{equation}
then
\begin{equation}
 g^{tt}=\frac{g_{\phi\phi}}{D},\qquad
 g^{t\phi}=-\frac{g_{t\phi}}{D},\qquad
 g^{\phi\phi}=\frac{g_{tt}}{D},
\end{equation}
while
\begin{equation}
 g^{rr}=\frac{\Delta}{(1+H_3)\Sigma},
 \qquad
 g^{\theta\theta}=\frac{1}{(1+H_5)\Sigma}.
\end{equation}
The angular expressions are evaluated using the exact transformation $x=\cos\theta$; the corresponding derivative obeys $d/d\theta=-\sin\theta\,d/dx$. This avoids numerical differencing of the deformation functions.

The inner termination surface is the Kerr horizon used by the supplied implementation,
\begin{equation}
 r_+=1+\sqrt{1-\chi^2},
\end{equation}
with the dimensionless normalization adopted above \cite{Cano:2019ore}. In the model considered here, the deformation parameters do not move this capture surface. Consequently, varying the cubic couplings changes the photon propagation outside the inner boundary while leaving the capture prescription fixed.

\section{Shadow construction}\label{sec:shadow}
\subsection{Hamiltonian formulation}
The null geodesics are evolved in Hamiltonian form.
The Hamiltonian is
\begin{equation}
	H=\frac12 g^{\mu\nu}p_\mu p_\nu,
\end{equation}
with the null condition $H=0$. 
Stationarity and axisymmetry imply two conserved covariant momenta,
\begin{equation}
 p_t=-E,\qquad p_\phi=L_z,
\end{equation}
where $E$ is the photon energy associated with the timelike Killing vector and $L_z$ is the axial angular momentum. The dynamical variables are therefore
\begin{equation}
 y=(r,\theta,p_r,p_\theta).
\end{equation}
 The evolution equations used by the integrator are
\begin{equation}
 \dot r=g^{rr}p_r,\qquad
 \dot\theta=g^{\theta\theta}p_\theta,
\end{equation}
and
\begin{equation}
 \dot p_r=-\frac12\,\partial_r g^{\mu\nu}p_\mu p_\nu,
 \qquad
 \dot p_\theta=-\frac12\,\partial_\theta g^{\mu\nu}p_\mu p_\nu.
\end{equation}
The derivatives of the inverse metric are evaluated analytically from the metric functions. Thus the ray-tracing calculation does not introduce finite-difference derivatives of the metric into the geodesic equations.

\subsection{Initial conditions}

The photon direction is defined in a local zero-angular-momentum-observer (ZAMO) tetrad. 
Let $(\alpha,\beta)$ denote the two screen coordinates. In Appendix~\ref{tetrad}, we introduce the tetrad and how screen coordinates are related to photon momentum. The finite-distance gnomonic convention used in the calculation defines the local spatial direction through
\begin{equation}
 X=-\frac{\alpha}{r_{\rm obs}},
 \qquad
 Y=\frac{\beta}{r_{\rm obs}},
 \qquad
 f=\frac{1}{\sqrt{1+X^2+Y^2}},
\end{equation}
so that the locally measured direction cosines are  given by
\begin{equation}
 n_r=-f,\qquad n_\phi=Xf,\qquad n_\theta=Yf.
\end{equation}
The normalization of the photon momentum is chosen so that $E=1$. The resulting covariant initial momenta are then obtained from the local tetrad and the metric. This procedure is repeated for every screen ray.

\subsection{Adaptive geodesic integration}
The geodesic system is integrated with the embedded eighth-order Dormand--Prince DOP853 method~\cite{DormandPrince1980}. For an attempted step, the code constructs the eighth-order solution and an embedded lower-order estimate and forms the standard DOP853 blended error measure. For each dynamical component $y_q$, the scale entering the error estimate is
\begin{equation}
 s_q={\rm atol}+{\rm rtol}\max\left(|y_q|,|y_q^{\rm new}|\right).
\end{equation}
The embedded error coefficients are combined into the DOP853 error estimator, and a step is accepted only when the resulting error measure does not exceed unity.

The production calculation uses
\begin{equation}
 {\rm rtol}=10^{-9},
 \qquad
 {\rm atol}=10^{-11},
\end{equation}
with nominal initial step $h_0=0.20$ and bounds
\begin{equation}
 10^{-7}\le h\le0.75.
\end{equation}
The adaptive controller changes the step size after accepted and rejected steps. In addition, if a ray cannot be advanced successfully, the calculation retries it with the fixed initial-step sequence
\begin{equation}
 h_0\times(1,0.25,0.05,0.01,0.002,0.0001).
\end{equation}
The retry sequence is not used to define the physical solution; it is a recovery mechanism for difficult trajectories.

For each initial screen direction the geodesic is followed until one of the termination conditions is reached. A ray is classified as captured when it reaches the prescribed inner boundary and as escaping when it reaches the outer radius.
A numerical failure is retained as a separate status and is not silently counted as either capture or escape. This distinction is important near the critical boundary, where the trajectory can spend a long affine-time interval close to the unstable photon region.

The coarse screen contains $N=220$ samples in each Cartesian screen direction. It is used to determine the approximate location of the captured region and to obtain a first estimate of its horizontal center. The actual shadow boundary is then constructed independently of the coarse-grid edge by probing the screen along polar directions.

\subsection{Boundary-root localization}
Let $\varphi$ be the polar angle on the observer's screen and let $R(\varphi)$ denote the radial distance from the adopted screen center. Along each fixed $\varphi$, the code searches for a pair of radii with different capture/escape classifications. Once such a bracket is found, the boundary radius is determined by one-dimensional bisection.
The production calculation uses $ N_{\rm bisect}=13$. 
The boundary calculation uses $N_{\rm angle}=1000$ polar directions. Thus the final curve is not obtained by fitting a small number of critical points: it is assembled from a large set of independently classified radial boundary roots.

\section{Numerical accuracy and convergence}\label{sec:numerics}
The code performs a series of consistency checks to validate the numerical integration of null geodesics. Initially, it verifies that the Hamiltonian and null constraints are satisfied within prescribed numerical tolerances, $ max |H|<10^{-3}$. During integration, it monitors deviations in the Hamiltonian and null condition to assess constraint preservation. It also checks the conservation of energy and angular momentum, represented by $p_t=-E$ and $p_\phi=L_z$, respectively, $max |p_t+E|  <10^{-5}$ and $max |p_\phi-L_z|<10^{-5}$. These checks help identify numerical errors and assess the reliability of the computed trajectories. The code also monitors the inverse-metric identity and tetrad orthonormality.

The numerical error analysis is designed to answer a concrete question: when the reported shadow changes, is the change physical or numerical? The calculation contains two other distinct numerical layers. The first determines the trajectory for a given initial photon direction. The second determines where the capture--escape transition occurs on the screen. We therefore vary the control parameters of these layers separately.

\subsection{Convergence of the geodesic integration}
The DOP853 convergence test varies the relative tolerance over
\begin{equation}
 {\rm rtol}=10^{-7},\;3\times10^{-8},\;10^{-8},\;3\times10^{-9},\;10^{-9},\;3\times10^{-10},
\end{equation}
while setting
\begin{equation}
 {\rm atol}=10^{-2}{\rm rtol}.
\end{equation}
All other numerical quantities are kept fixed: the screen resolution, number of boundary angles, bisection depth, observer position, metric parameters, step-size bounds, capture and escape surfaces, and retry sequence. The finest tolerance is used only as a numerical reference; it is not assumed to be an exact solution.

For each tolerance the complete boundary construction is repeated. We compare the shadow area. Because the same boundary-root procedure is retained at every tolerance, this test isolates the ODE integration error rather than mixing it with root-finding error.

\subsection{Convergence of the boundary root}
A separate test varies only the number of bisection iterations,
\begin{equation}
 N_{\rm bisect}=8,\;10,\;12,\;13,\;15,\;17,\;20,
\end{equation}
while the DOP853 tolerances remain fixed at their production values. The coarse captured mask and its resulting screen center are determined once and then held fixed throughout this test. For each bisection depth, every boundary ray is nevertheless reclassified and its radial root is recomputed. The 20-iteration boundary is used as the reference.

\begin{figure}[h!]
		\centering
		\includegraphics[width=10cm]{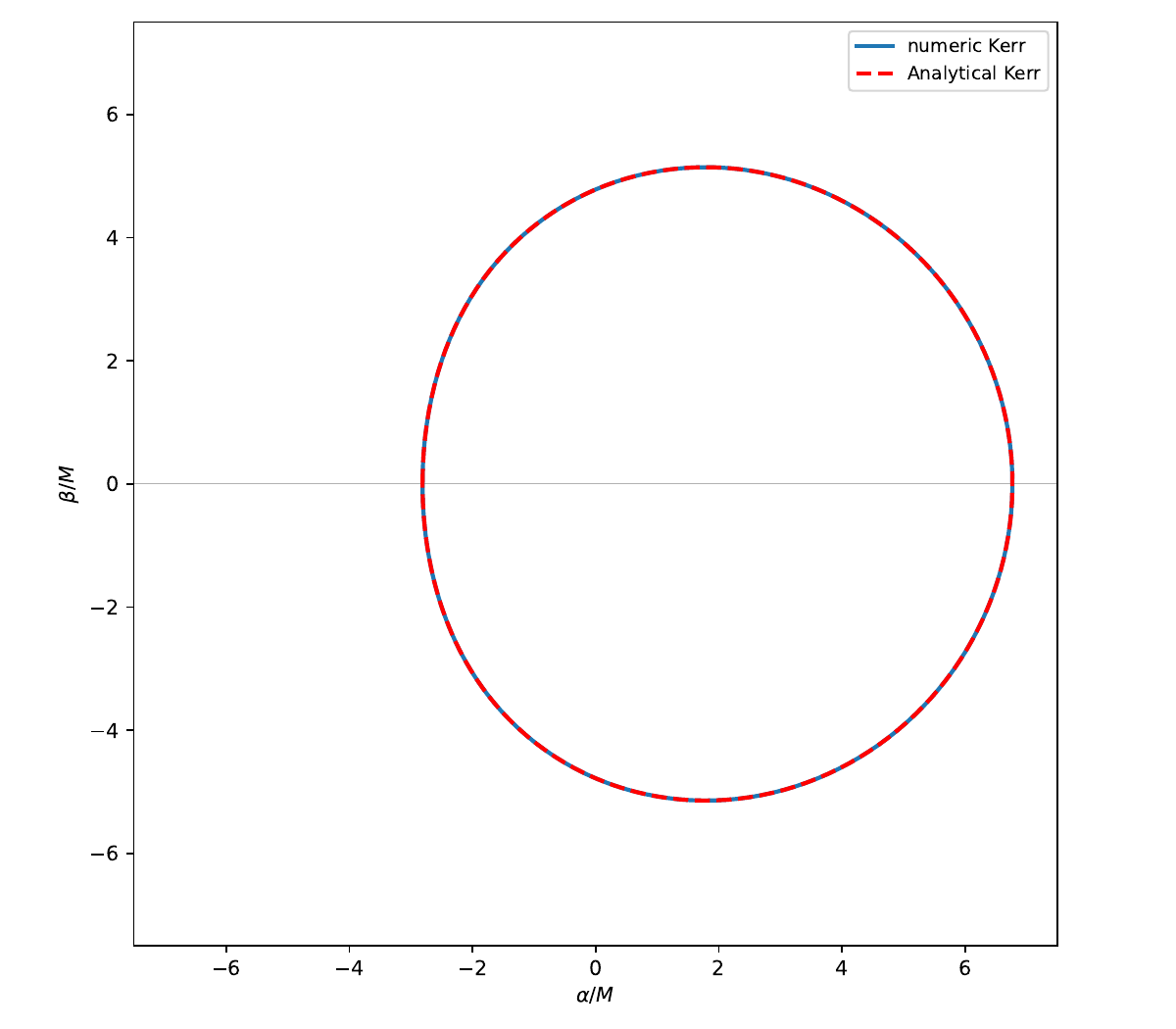}
		\caption{Analytical Kerr shadow and numerical shadow for $a/M=0.9$ and observer angle $\theta_o=\pi/2$.}
		\label{fig0}
\end{figure}

\begin{figure}[h!]
		\centering
		\includegraphics[width=7cm]{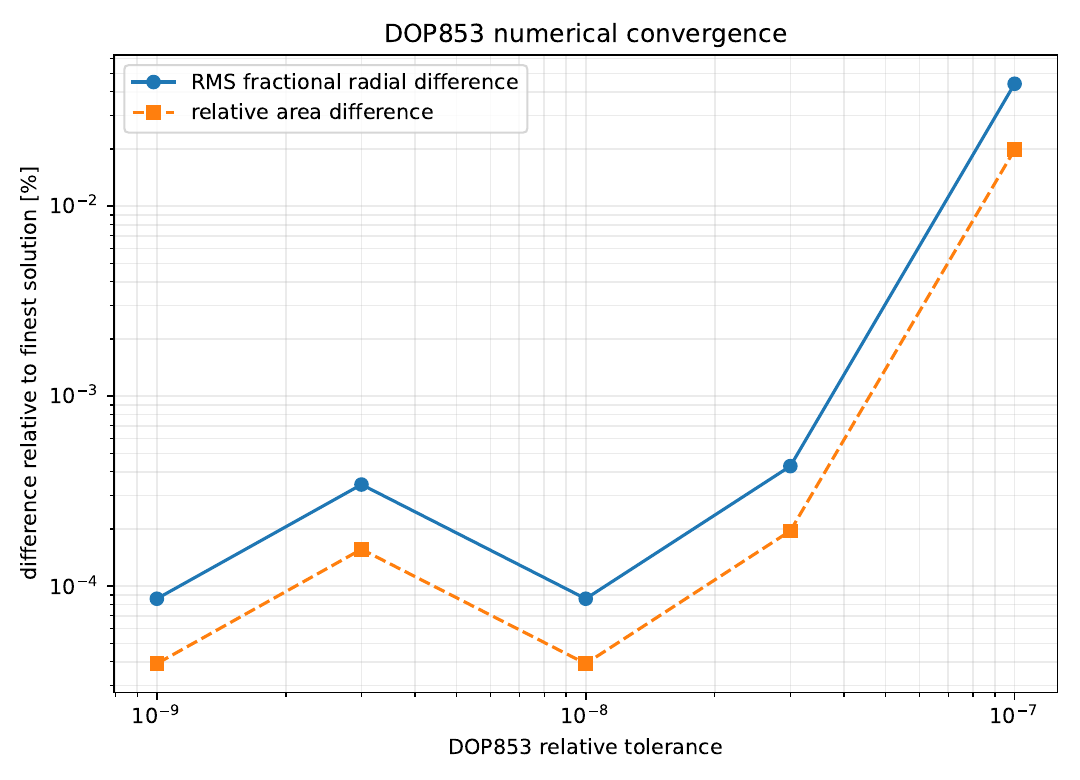}
		\includegraphics[width=7cm]{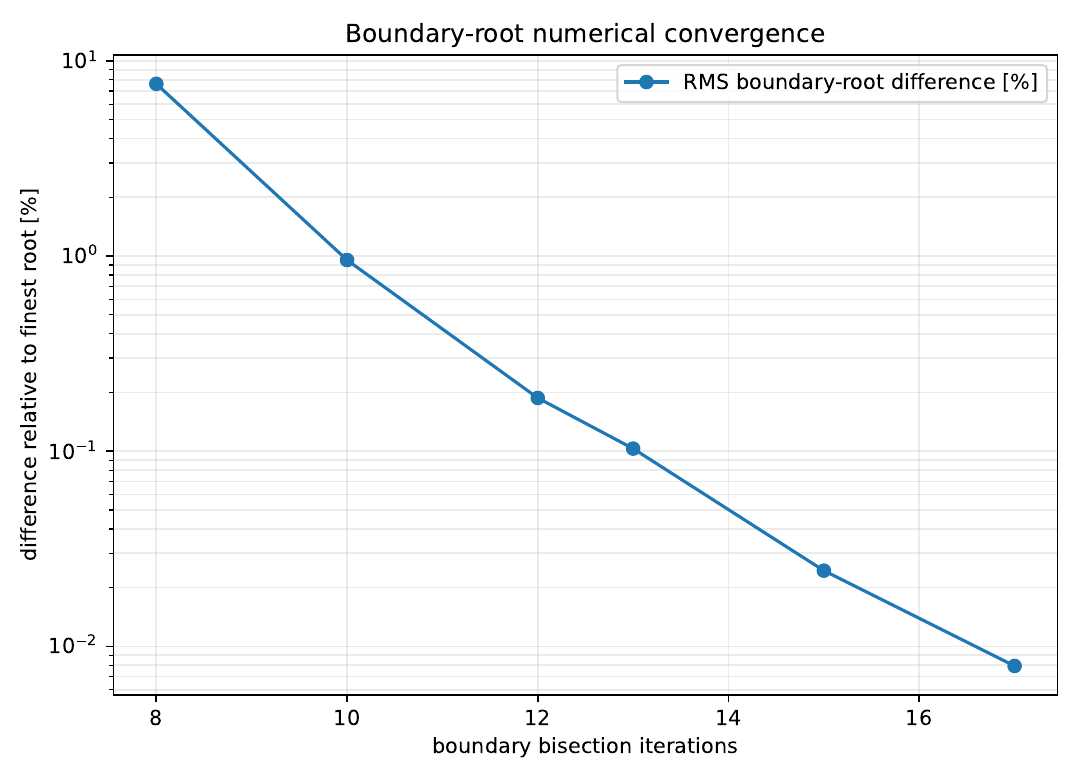}
		\caption{Convergence tests. Left: Relative area difference and root mean square of radial difference as a function of tolerances. Right: convergence of the algorithm for finding the shadow boundary.}
		\label{fig1-2}
\end{figure}

The comparison includes the relative change in area. The nominal root bracket decreases geometrically with the number of iterations, providing a direct numerical expectation for convergence. In Figure~\ref{fig0}, we plot our result and compare with analytic Kerr shadow. We used $a/M=0.9$ and $\theta_o=\pi/2$. In Figure~\ref{fig1-2}, we show the convergence of our code by changing tolerances and boundary iterations, respectively.

The two tests are complementary. Stability with respect to DOP853 tolerance but sensitivity to bisection depth would identify boundary localization as the limiting error. Stability with respect to bisection but sensitivity to the integration tolerance would instead indicate that the geodesic solver controls the error. Convergence under both refinements is the criterion adopted here for the production calculation.

\section{Dependence on the first cubic deformation $\beta_5$}\label{sec:b5}
The first deformation parameter is $\beta_5$. At fixed spin, observer position, inclination, and second deformation parameter, the entire shadow calculation is repeated as $\beta_5$ is varied. The metric functions, their analytic derivatives, the initial photon momenta, and all geodesic trajectories are recomputed for every parameter value.

\begin{figure}[ht!]
	\centering
	\includegraphics[width=10cm]{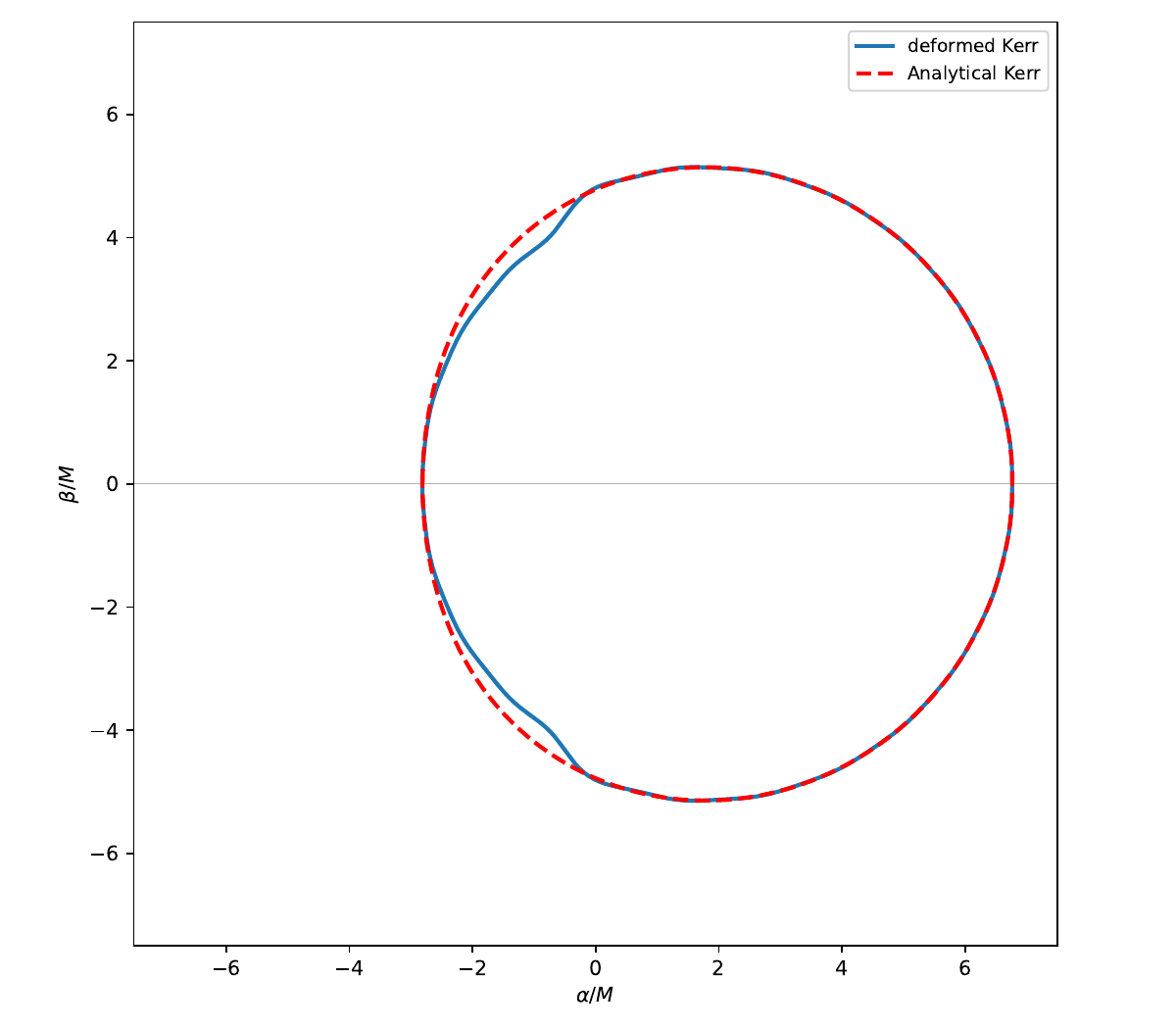}
	\caption{Deformed Kerr shadow for $\beta_5=0.01$, $a/M=0.9$, $\beta_6=0$ and observer inclination $\theta_o=\pi/2$.}
	\label{figb5}
\end{figure}

\begin{figure}[h!]
	\centering
	\includegraphics[width=7cm]{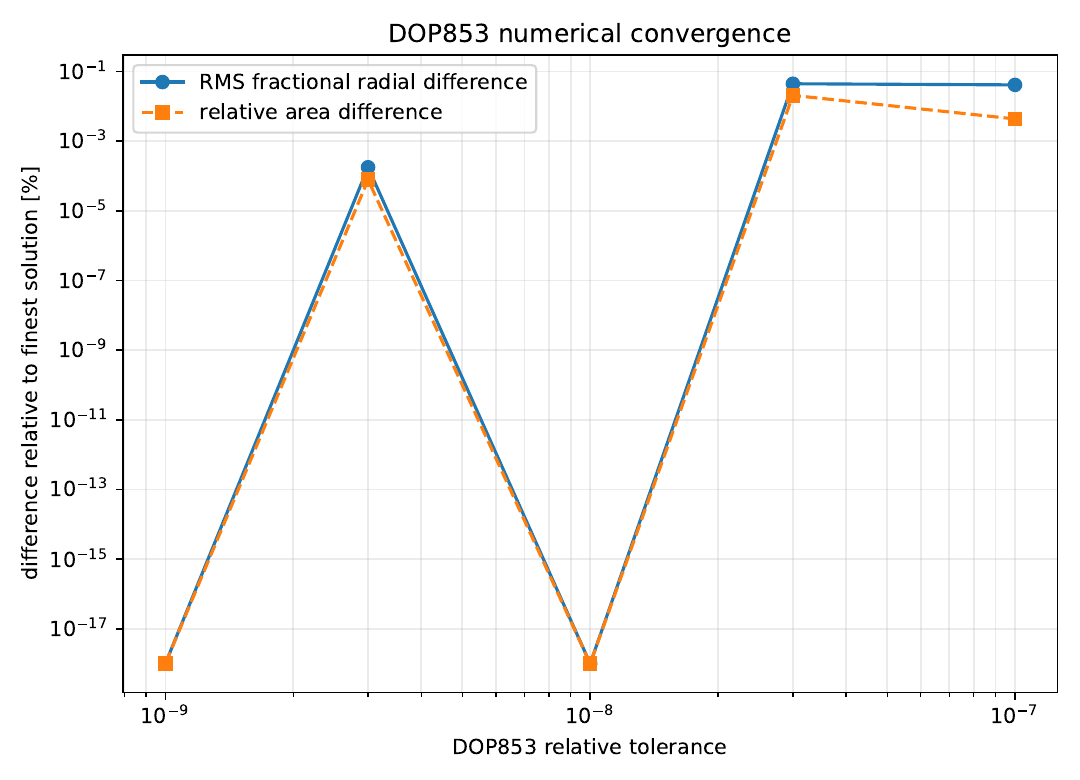}
	\includegraphics[width=7cm]{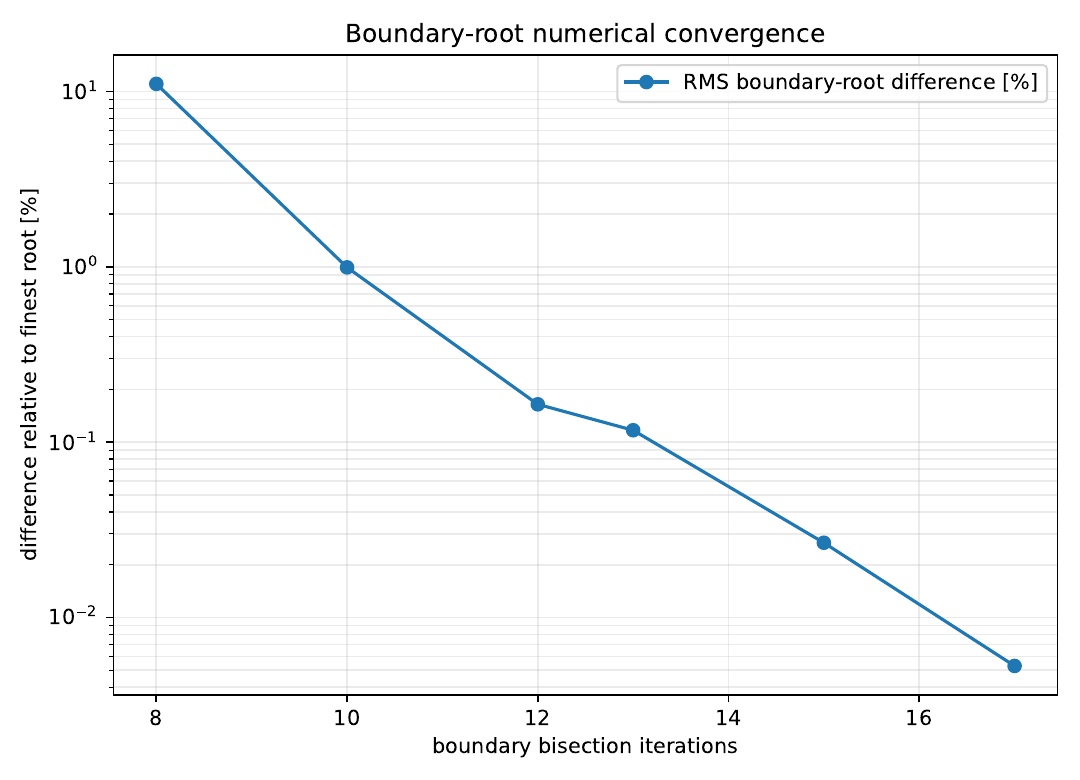}
	\caption{Convergence tests for figure~\ref{figb5}}
	\label{figb5-1}
\end{figure}

The numerical comparison is performed against the undeformed Kerr sector.
The same numerical resolution and convergence-controlled integration prescription are retained across the scan. Thus a systematic trend with $\beta_5$ is not generated by changing the numerical accuracy from one parameter value to another. The admissible parameter range is additionally restricted by the requirement that the exterior metric remain regular and Lorentzian throughout the region sampled by the ray tracer. The result is presented in Figure~\ref{figb5}. Convergence is shown in Figure~\ref{figb5-1}.

\section{Dependence on the second cubic deformation $\beta_6$}\label{sec:b6}
The second deformation parameter is $\beta_6$. Its influence is studied with the same procedure used for $\beta_5$. For each value, the exact deformation functions are evaluated, the finite-distance photon initial data are reconstructed, and the null geodesics are integrated until capture or escape.

\begin{figure}[h!]
	\centering
	\includegraphics[width=12cm]{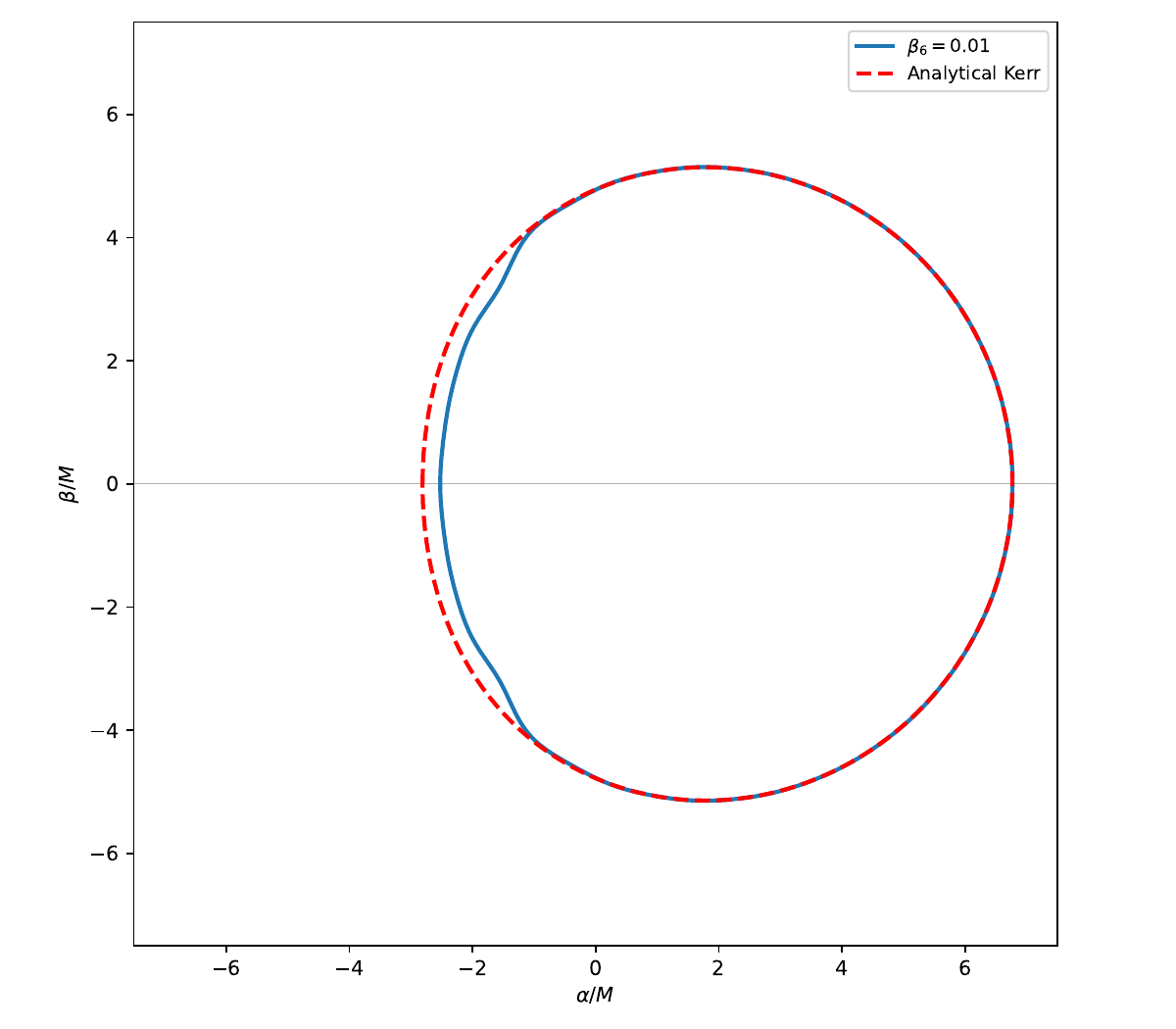}
	\caption{Deformed Kerr shadow for $\beta_6=0.01$, $a/M=0.9$, $\beta_5=0$ and observer inclination $\theta_o=\pi/2$.}
	\label{fig3}
\end{figure}

\begin{figure}[h!]
	\centering
	\includegraphics[width=7cm]{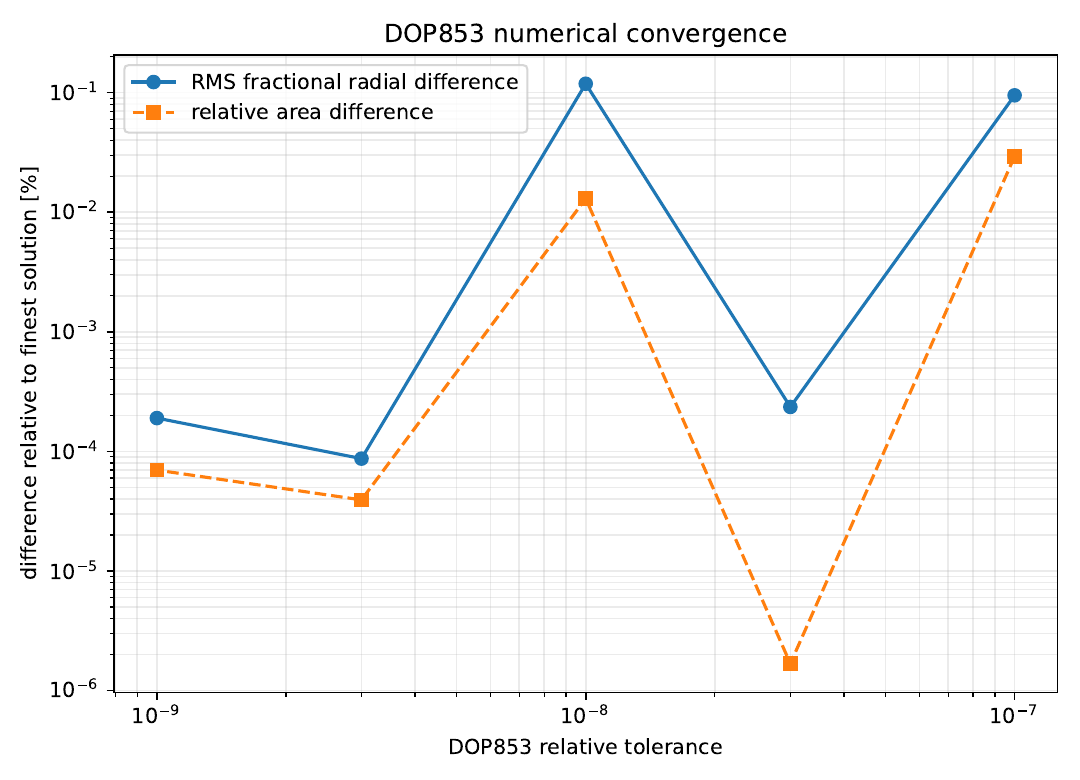}
		\includegraphics[width=7cm]{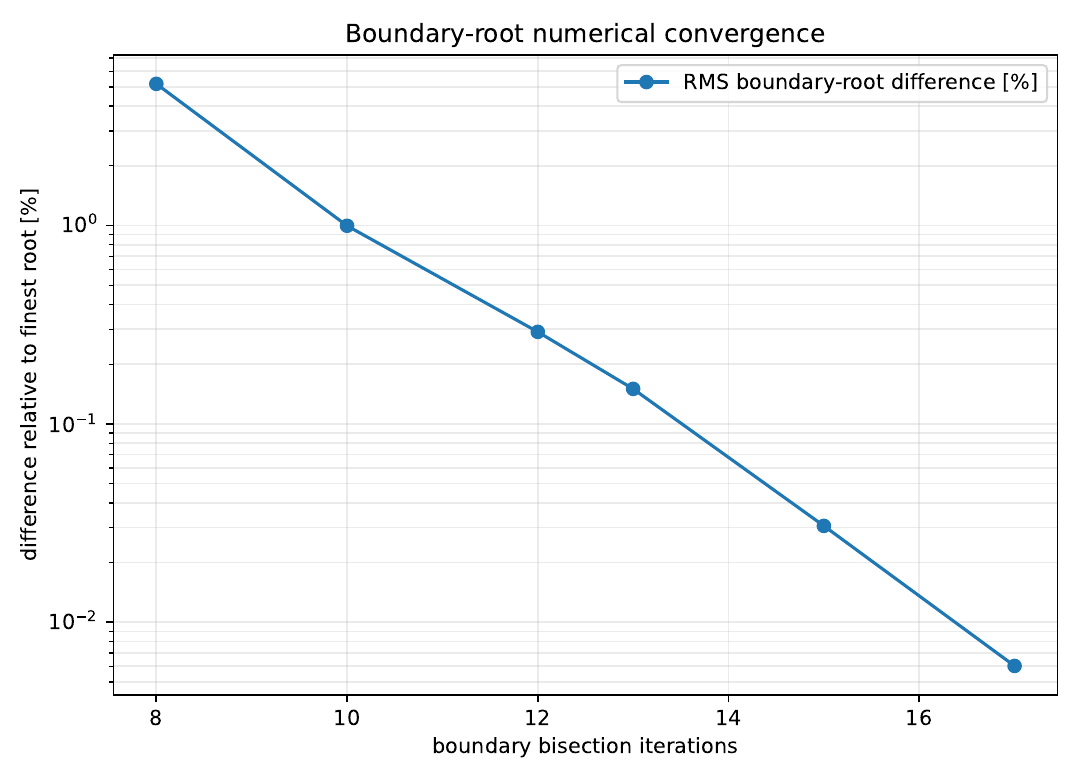}
	\caption{Convergence tests for Figure~\ref{fig3} }
	\label{fig45}
\end{figure}

\begin{figure}[h!]
	\centering
	\includegraphics[width=12cm]{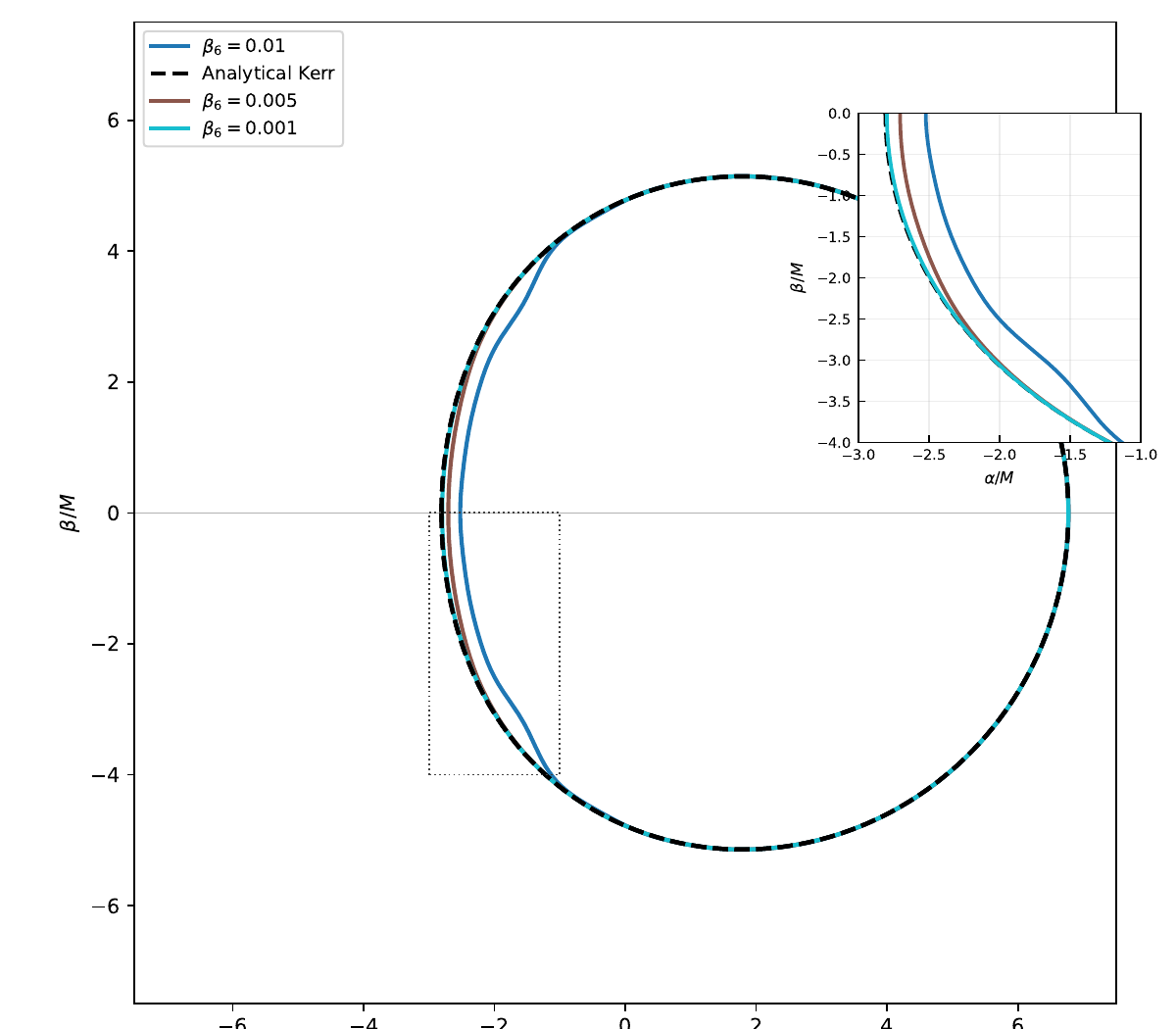}
	\caption{Deformed Kerr shadow for various $\beta_6$. We have  $a/M=0.9$, $\beta_5=0$ and observer inclination $\theta_o=\pi/2$.}
	\label{fig6}
\end{figure}

\begin{figure}[h!]
	\centering
	\includegraphics[width=12cm]{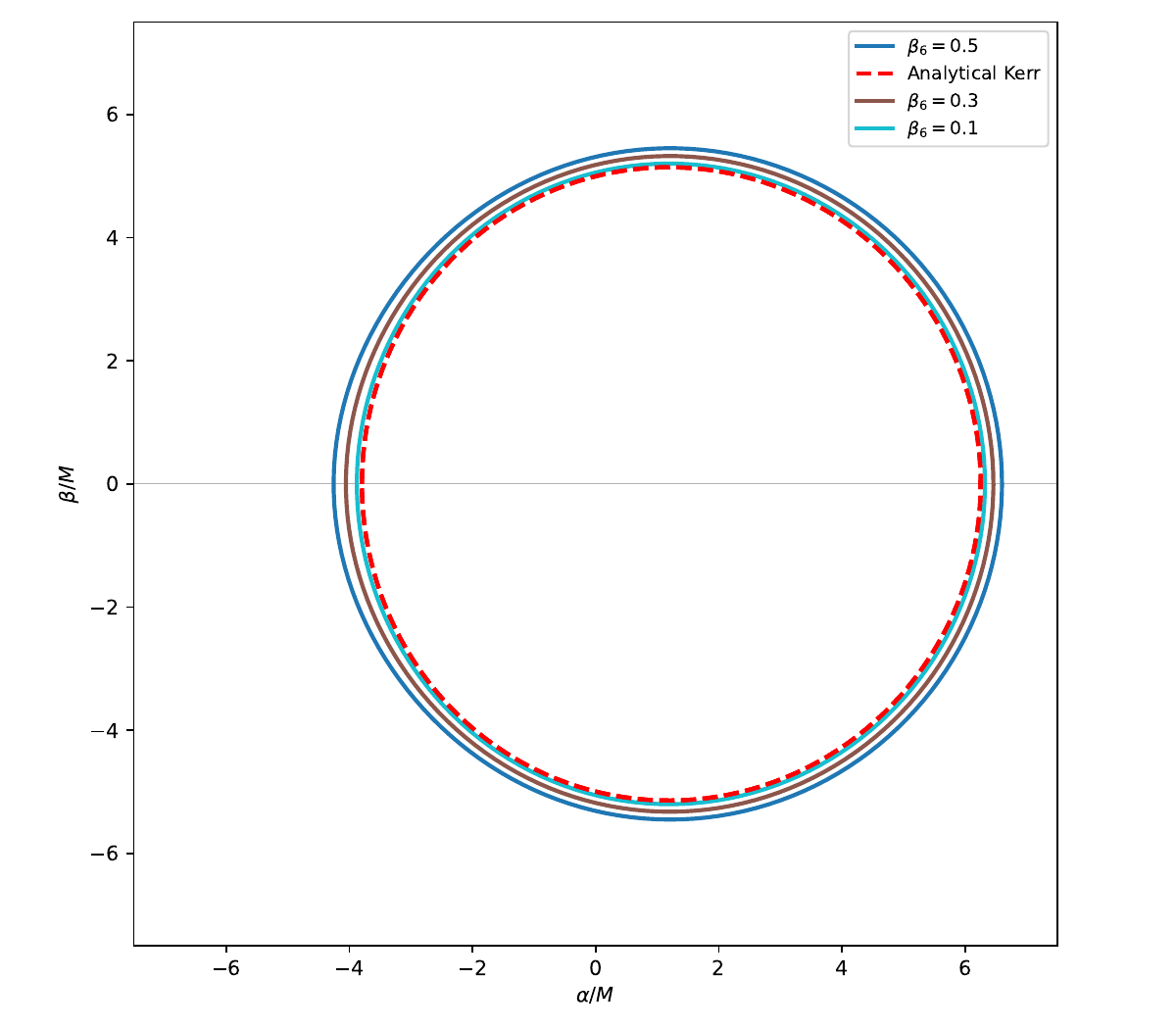}
	\caption{Deformed Kerr shadow for various $\beta_6$. We have  $a/M=0.6$, $\beta_5=0$ and observer inclination $\theta_o=\pi/2$.}
	\label{fig7}
\end{figure}

\begin{figure}[h!]
	\centering
	\includegraphics[width=12cm]{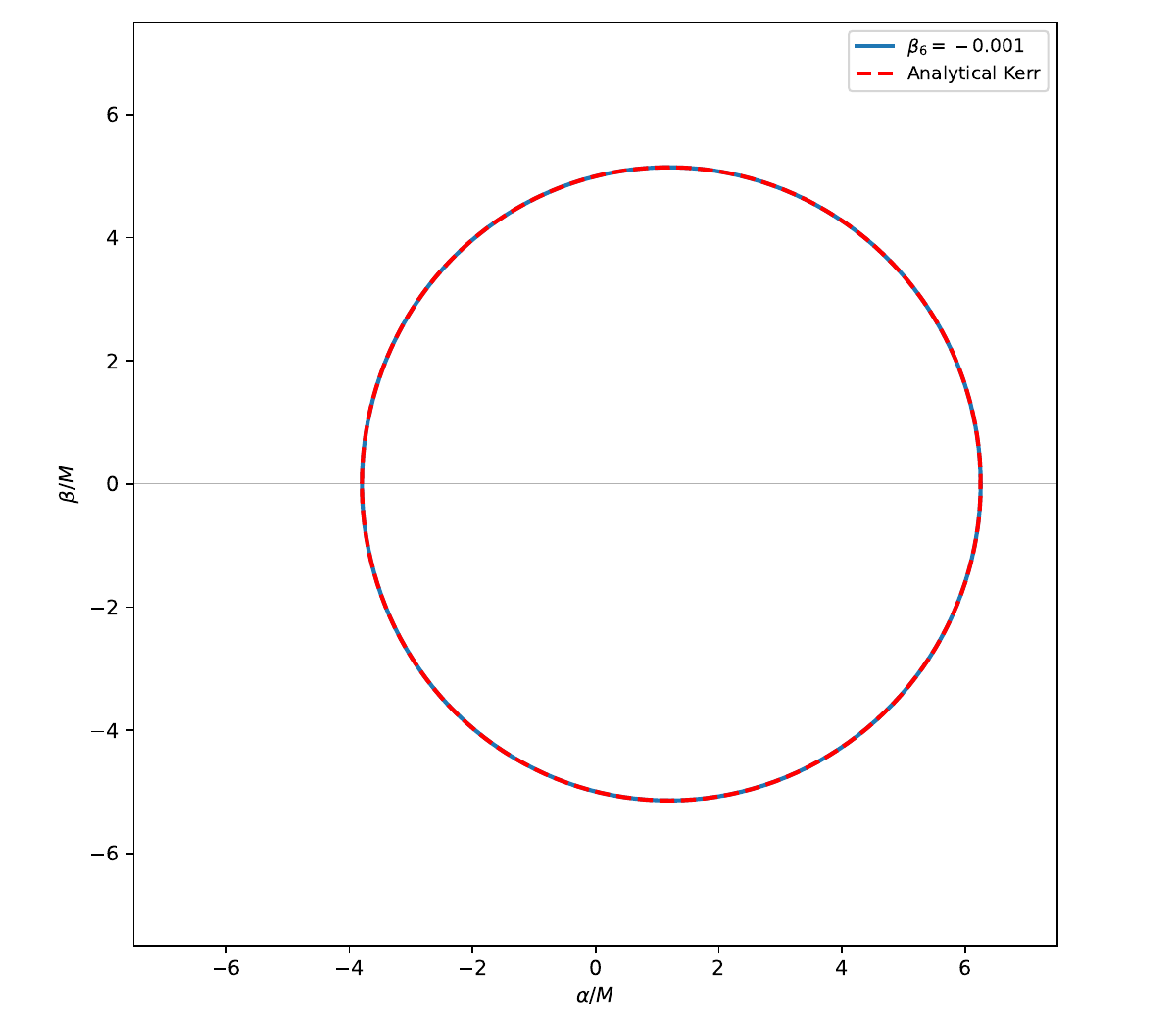}
	\caption{Deformed Kerr shadow for  negative $\beta_6=-0.001$. We have  $a/M=0.6$, $\beta_5=0$ and observer inclination $\theta_o=\pi/2$.}
	\label{fig8}
\end{figure}

\begin{figure}[h!]
	\centering
	\includegraphics[width=7cm]{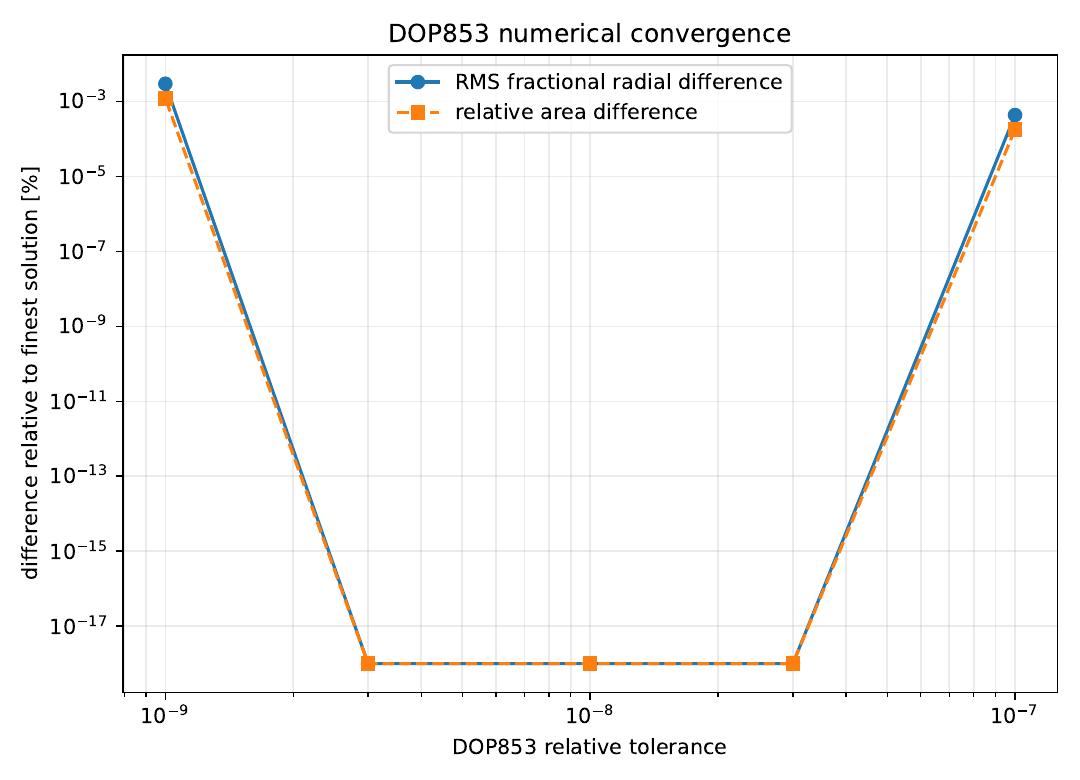}
	\includegraphics[width=7cm]{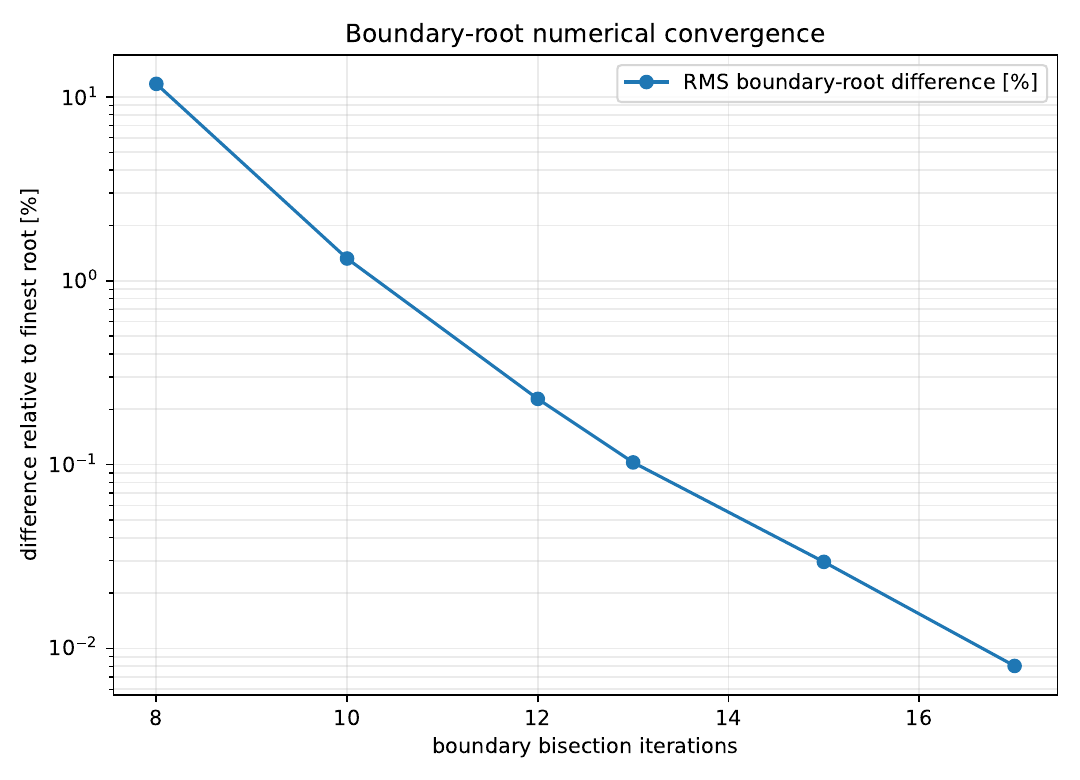}
	\caption{Convergence tests for Figure~\ref{fig8} }
	\label{fig910}
\end{figure}

The resulting boundaries are compared through their area differences. 
The numerical treatment is deliberately identical throughout the scan. This is particularly relevant when the deformation is small: a physical effect can be smaller than a numerical artifact if the error controls are allowed to vary between parameter values. The independent convergence tests therefore provide the baseline against which the parameter-dependent changes are interpreted. In Figure~\ref{fig3}, we show deformed Kerr shadow for $\beta_6=0.01$, $a/M=0.9$, $\beta_5=0$ and observer inclination $\theta_o=\pi/2$. The convergence tests are shown in Figure~\ref{fig45}. To compare the effect of $\beta_6$, we show shadow for various parameters in Figures~\ref{fig6} and~\ref{fig7}. For negative $\beta_6$, we find it difficult to identify parameter ranges in which the shadow deviates noticeably from that of Kerr. We plot our result in Figure~\ref{fig8} and convergence in Figure~\ref{fig910}.

\section{Dependence on observer inclination}\label{sec:inclination}
The observer inclination is denoted by $\theta_{\rm o}$. It enters the construction through the location of the observer in the angular direction and through the orientation of the local orthonormal frame used to initialize photon momenta.

\begin{figure}[h!]
	\centering
	\includegraphics[width=12cm]{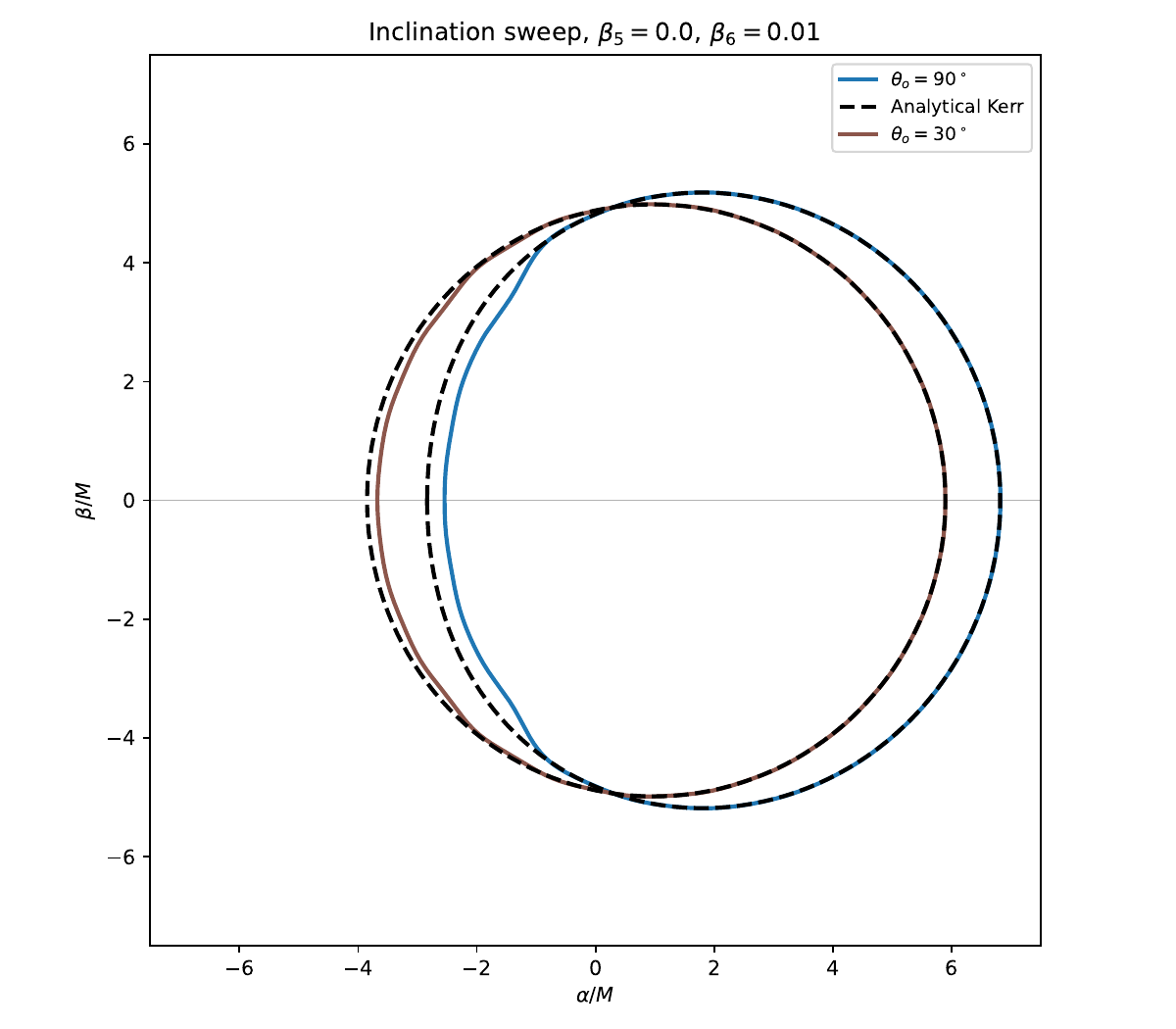}
	\caption{Deformed Kerr shadow for various $\theta_o$. We have $\beta_6=0.01$, $a/M=0.9$, $\beta_5=0$.}
	\label{figthet}
\end{figure}

At finite observer radius, changing $\theta_{\rm o}$ changes the local mapping between photon directions and screen coordinates. The inclination scan therefore represents a direct change in viewing geometry rather than a post-processing transformation of a single asymptotic critical curve. For every inclination, the capture/escape boundary is recomputed from the geodesics. In Figure~\ref{figthet}, we plot the effect of observer  inclination on the shadow and compare it with the Kerr shadow for each inclination.

\section{Discussion and conclusion }\label{sec:discussion}
We have presented a direct numerical construction of Kerr-cubic black-hole shadows based on finite-distance ray tracing. The method starts from the exact supplied metric functions, constructs photon initial data in a local ZAMO frame, integrates the null Hamilton equations with adaptive DOP853, classifies rays by their capture or escape outcome, and determines the critical screen boundary by one-dimensional bisection.

A central feature of the calculation is the explicit separation of numerical error sources. The DOP853 tolerance study tests the convergence of the individual geodesics, while the independent boundary-root study tests the convergence of the observable boundary itself. The two tests use fixed settings for the other numerical layer, making their interpretation straightforward. Constraint and tetrad diagnostics provide an additional consistency check on the implementation.

The use of an embedded adaptive eighth-order integrator controls the local error accumulated along each ray, while the fixed retry ladder provides a reproducible recovery mechanism for trajectories that are difficult to advance. The retry sequence is held fixed in the convergence studies, so tightening the requested tolerance changes the integration accuracy without simultaneously changing the recovery algorithm. The analytic derivatives of the inverse metric further remove finite-difference noise from the Hamilton equations.

The boundary calculation is equally explicit. A large set of polar directions is used, each boundary radius is bracketed by actual capture and escape classifications, and the root is then refined by bisection. This is preferable to estimating the critical curve from a small set of sampled rays and subsequently interpolating it, because the convergence parameter has a direct interpretation: each additional bisection step halves the radial bracket associated with that boundary ray.

The supplied implementation also performs metric, tetrad, and Hamiltonian consistency checks. These diagnostics are not substitutes for convergence, but they provide useful independent tests of the numerical implementation. In combination with the two refinement studies, they give a layered validation strategy: algebraic consistency of the metric and frame, controlled integration of individual rays, and controlled localization of the final boundary.

\section*{Acknowledgements}
The work of YP is supported by the National Natural Science Foundation of China (NSFC) under Grant No. 12575076 and No. 12247103 and by the National Key R$\&$D Program No. 2022YFE0134300. The work of SM was supported in part by JSPS (Japan Society for the Promotion of Science) KAKENHI Grant No.\ JP24K07017 and World Premier International Research Center Initiative (WPI), MEXT, Japan.

\appendix
\section{Observer tetrad}
\label{tetrad}

A ZAMO has vanishing angular momentum per unit photon-independent
observer normalization, $u_\phi=0$, where $u^\mu$ is the observer's
four-velocity. Its angular velocity relative to the stationary
coordinate frame is
\begin{equation}
	\omega=-\frac{g_{t\phi}}{g_{\phi\phi}}.
	\label{eq:zamo_omega}
\end{equation}
The lapse function is defined by
\begin{equation}
	N=\frac{1}{\sqrt{-g^{tt}}},
	\label{eq:zamo_lapse}
\end{equation}
where $g^{tt}$ is the $tt$ component of the inverse metric.
The ZAMO four-velocity is therefore
\begin{equation}
	u^\mu=\frac{1}{N}
	\left(1,0,0,\omega\right),
	\qquad
	u^\mu u_\mu=-1.
	\label{eq:zamo_four_velocity}
\end{equation}
The angular velocity $\omega$ accounts for frame dragging:
although the observer has zero angular momentum, the observer
is generally not stationary with respect to the coordinates.

We construct an orthonormal tetrad
$\{e_{(t)}^\mu,e_{(r)}^\mu,e_{(\theta)}^\mu,e_{(\phi)}^\mu\}$
whose timelike member is the observer's four-velocity,
$e_{(t)}^\mu=u^\mu$. The spatial basis vectors are
\begin{align}
	e_{(r)}^\mu
	&=\left(0,\frac{1}{\sqrt{g_{rr}}},0,0\right),
	\label{eq:tetrad_r}\\
	e_{(\theta)}^\mu
	&=\left(0,0,\frac{1}{\sqrt{g_{\theta\theta}}},0\right),
	\label{eq:tetrad_theta}\\
	e_{(\phi)}^\mu
	&=\left(0,0,0,\frac{1}{\sqrt{g_{\phi\phi}}}\right).
	\label{eq:tetrad_phi}
\end{align}
These expressions apply in the exterior region where the indicated
spatial metric components are positive. The tetrad satisfies
\begin{equation}
	g_{\mu\nu}e_{(a)}^\mu e_{(b)}^\nu
	=\eta_{(a)(b)},
	\qquad
	\eta_{(a)(b)}
	=\operatorname{diag}(-1,1,1,1).
	\label{eq:tetrad_orthonormality}
\end{equation}
In particular, the azimuthal basis vector is orthogonal to the
ZAMO four-velocity because
$g_{t\phi}+\omega g_{\phi\phi}=0$.

\paragraph{Projection from the screen to local directions.}

The screen coordinates $(\alpha,\beta)$ specify the apparent
position of a ray on the observer's image plane. In this work,
we adopt a finite-distance gnomonic screen convention. Define
the dimensionless screen coordinates
\begin{equation}
	X=-\frac{\alpha}{r_{\rm obs}},
	\qquad
	Y=\frac{\beta}{r_{\rm obs}}.
	\label{eq:screen_xy}
\end{equation}
The associated local spatial direction is normalized using
\begin{equation}
	f=\frac{1}{\sqrt{1+X^2+Y^2}}.
	\label{eq:screen_normalization}
\end{equation}
With the sign convention that an initially inward-directed ray
has a negative radial component, the direction cosines measured
in the ZAMO frame are
\begin{equation}
	n_{(r)}=-f,\qquad
	n_{(\phi)}=Xf,\qquad
	n_{(\theta)}=Yf.
	\label{eq:screen_direction}
\end{equation}
They obey
\begin{equation}
	n_{(r)}^2+n_{(\theta)}^2+n_{(\phi)}^2=1,
	\label{eq:direction_normalization}
\end{equation}
so the initial photon direction is a unit spatial vector in the
local orthonormal frame. Equations~\eqref{eq:screen_xy}--%
\eqref{eq:screen_direction} specify the screen convention used
in our numerical implementation; they should not be confused
with the asymptotic celestial-coordinate formulae commonly used
for an observer at infinity.

\paragraph{From local photon momentum to coordinate momentum.}

Let $E_{\rm loc}>0$ denote the photon energy measured by the ZAMO.
The photon four-momentum can be written in the tetrad as
\begin{equation}
	p^\mu=E_{\rm loc}
	\left[
	u^\mu
	+n_{(r)}e_{(r)}^\mu
	+n_{(\theta)}e_{(\theta)}^\mu
	+n_{(\phi)}e_{(\phi)}^\mu
	\right].
	\label{eq:photon_tetrad_decomposition}
\end{equation}
This decomposition automatically enforces the local null condition
$p^\mu p_\mu=0$ because the spatial direction has unit norm.

Stationarity and axial symmetry imply two conserved quantities:
\begin{equation}
	E=-p_t,\qquad L_z=p_\phi.
	\label{eq:conserved_quantities}
\end{equation}
We set the energy at infinity to $E=1$, which fixes the overall
normalization of the photon momentum without affecting its
trajectory. Using
$u^\mu=N^{-1}(1,0,0,\omega)$, the covariant time component is
\begin{equation}
	p_t=E_{\rm loc}
	\left(
	-N+n_{(\phi)}
	\frac{g_{t\phi}}{\sqrt{g_{\phi\phi}}}
	\right).
	\label{eq:pt_zamo}
\end{equation}
Consequently, the local photon energy required to impose $-p_t=E$
is
\begin{equation}
	E_{\rm loc}
	=
	\frac{E}{
		N-n_{(\phi)}g_{t\phi}/\sqrt{g_{\phi\phi}}
	}.
	\label{eq:local_energy}
\end{equation}
The contravariant coordinate components then follow directly:
\begin{align}
	p^t&=\frac{E_{\rm loc}}{N},
	\label{eq:pt_contravariant}\\
	p^\phi&=\frac{\omega E_{\rm loc}}{N}
	+\frac{E_{\rm loc}n_{(\phi)}}{\sqrt{g_{\phi\phi}}},
	\label{eq:pphi_contravariant}\\
	p^r&=\frac{E_{\rm loc}n_{(r)}}{\sqrt{g_{rr}}},
	\label{eq:pr_contravariant}\\
	p^\theta&=\frac{E_{\rm loc}n_{(\theta)}}
	{\sqrt{g_{\theta\theta}}}.
	\label{eq:ptheta_contravariant}
\end{align}
Lowering the index with the metric gives
\begin{equation}
	p_\mu=g_{\mu\nu}p^\nu.
	\label{eq:momentum_lowering}
\end{equation}
In particular, this yields the initial values of $p_r$,
$p_\theta$, and $L_z=p_\phi$ used by the Hamiltonian ray tracer.
The conserved value $p_t=-E$ is imposed by the normalization above.

\clearpage

\bibliographystyle{JHEP}
\bibliography{ref}

@article{Fernandes:2025vxg,
    author = "Fernandes, Pedro G. S.",
    title = "{Leading Effective Field Theory Corrections to the Kerr Metric at All Spins}",
    eprint = "2512.02338",
    archivePrefix = "arXiv",
    primaryClass = "gr-qc",
    doi = "10.1103/5rk7-zljb",
    journal = "Phys. Rev. Lett.",
    volume = "136",
    number = "24",
    pages = "241401",
    year = "2026"
}

@article{Ozkan:2024euj,
    author = "Ozkan, Mehmet and Pang, Yi and Sezgin, Ergin",
    title = "{Higher derivative supergravities in diverse dimensions}",
    eprint = "2401.08945",
    archivePrefix = "arXiv",
    primaryClass = "hep-th",
    reportNumber = "MI-HET-826",
    doi = "10.1016/j.physrep.2024.07.002",
    journal = "Phys. Rept.",
    volume = "1086",
    pages = "1--95",
    year = "2024"
}

@article{Goroff:1985th,
    author = "Goroff, Marc H. and Sagnotti, Augusto",
    title = "{The Ultraviolet Behavior of Einstein Gravity}",
    reportNumber = "CALT-68-1289, LBL-19995, UCB-PTH-85-34",
    doi = "10.1016/0550-3213(86)90193-8",
    journal = "Nucl. Phys. B",
    volume = "266",
    pages = "709--736",
    year = "1986"
}

@article{Metsaev:1986yb,
    author = "Metsaev, R. R. and Tseytlin, Arkady A.",
    title = "{Curvature Cubed Terms in String Theory Effective Actions}",
    reportNumber = "Print-86-1366 (LEBEDEV INST), LEBEDEV-86-316",
    doi = "10.1016/0370-2693(87)91527-9",
    journal = "Phys. Lett. B",
    volume = "185",
    pages = "52--58",
    year = "1987"
}

@article{Allahyari:2025bvf,
	author = "Allahyari, Alireza and Ma, Liang and Mukohyama, Shinji and Pang, Yi",
	title = "{Constraining cubic curvature corrections to general relativity with quasi-periodic oscillations}",
	eprint = "2506.22548",
	archivePrefix = "arXiv",
	primaryClass = "gr-qc",
	reportNumber = "YITP-25-92, IPMU25-0031",
	doi = "10.1088/1475-7516/2025/11/011",
	journal = "JCAP",
	volume = "11",
	pages = "011",
	year = "2025"
}

@article{DormandPrince1980,
  author  = {Dormand, J. R. and Prince, P. J.},
  title   = {A family of embedded Runge-Kutta formulae},
  journal = {Journal of Computational and Applied Mathematics},
  volume  = {6},
  number  = {1},
  pages   = {19--26},
  year    = {1980},
  doi     = {10.1016/0771-050X(80)90013-3}
}

@article{eht1,author={{Event Horizon Telescope Collaboration}},title={First {M87} Event Horizon Telescope Results. {I}. The Shadow of the Supermassive Black Hole},journal={Astrophys. J. Lett.},volume={875},pages={L1},year={2019}}

@article{eht2,author={{Event Horizon Telescope Collaboration}},title={First {M87} Event Horizon Telescope Results. {II}. Array and Instrumentation},journal={Astrophys. J. Lett.},volume={875},pages={L2},year={2019}}

@article{eht3,author={{Event Horizon Telescope Collaboration}},title={First {M87} Event Horizon Telescope Results. {III}. Data Processing and Calibration},journal={Astrophys. J. Lett.},volume={875},pages={L3},year={2019}}

@article{eht4,author={{Event Horizon Telescope Collaboration}},title={First {M87} Event Horizon Telescope Results. {IV}. Imaging the Central Supermassive Black Hole},journal={Astrophys. J. Lett.},volume={875},pages={L4},year={2019}}

@article{eht5,author={{Event Horizon Telescope Collaboration}},title={First {M87} Event Horizon Telescope Results. {V}. Physical Origin of the Asymmetric Ring},journal={Astrophys. J. Lett.},volume={875},pages={L5},year={2019}}

@article{eht6,author={{Event Horizon Telescope Collaboration}},title={First {M87} Event Horizon Telescope Results. {VI}. The Shadow and Mass of the Central Black Hole},journal={Astrophys. J. Lett.},volume={875},pages={L6},year={2019}}

@article{eht7,author={{Event Horizon Telescope Collaboration}},title={First {M87} Event Horizon Telescope Results. {VII}. Polarization of the Ring},journal={Astrophys. J. Lett.},volume={910},pages={L12},year={2021}}

@article{eht8,author={{Event Horizon Telescope Collaboration}},title={First {M87} Event Horizon Telescope Results. {VIII}. Magnetic Field Structure near the Event Horizon},journal={Astrophys. J. Lett.},volume={910},pages={L13},year={2021}}

@article{sgra1,author={{Event Horizon Telescope Collaboration}},title={First {Sagittarius A*} Event Horizon Telescope Results. {I}. The Shadow of the Supermassive Black Hole in the Center of the {Milky Way}},journal={Astrophys. J. Lett.},volume={930},pages={L12},year={2022}}

@article{sgra2,author={{Event Horizon Telescope Collaboration}},title={First {Sagittarius A*} Event Horizon Telescope Results. {II}. {EHT} and Multiwavelength Observations, Data Processing, and Calibration},journal={Astrophys. J. Lett.},volume={930},pages={L13},year={2022}}

@article{sgra3,author={{Event Horizon Telescope Collaboration}},title={First {Sagittarius A*} Event Horizon Telescope Results. {III}. Imaging of the Galactic Center Supermassive Black Hole},journal={Astrophys. J. Lett.},volume={930},pages={L14},year={2022}}

@article{sgra4,author={{Event Horizon Telescope Collaboration}},title={First {Sagittarius A*} Event Horizon Telescope Results. {IV}. Variability, Morphology, and Black Hole Mass},journal={Astrophys. J. Lett.},volume={930},pages={L15},year={2022}}

@article{sgra5,author={{Event Horizon Telescope Collaboration}},title={First {Sagittarius A*} Event Horizon Telescope Results. {V}. Testing Astrophysical Models of the Galactic Center Black Hole},journal={Astrophys. J. Lett.},volume={930},pages={L16},year={2022}}

@article{sgra6,author={{Event Horizon Telescope Collaboration}},title={First {Sagittarius A*} Event Horizon Telescope Results. {VI}. Testing the Black Hole Metric},journal={Astrophys. J. Lett.},volume={930},pages={L17},year={2022}}

@article{eht2018,author={{Event Horizon Telescope Collaboration}},title={The persistent shadow of the supermassive black hole of {M87}. {I}. Observations, calibration, imaging, and analysis},journal={Astron. Astrophys.},volume={681},pages={A79},year={2024}}

@article{ehtcharge,author={Kocherlakota, Prashant and others},title={Constraints on black-hole charges with the 2017 {EHT} observations of {M87*}},journal={Phys. Rev. D},volume={103},pages={104047},year={2021}}

@article{oppenheimer,author={Oppenheimer, J. R. and Snyder, H.},title={On continued gravitational contraction},journal={Phys. Rev.},volume={56},pages={455--459},year={1939}}

@article{penrose65,author={Penrose, Roger},title={Gravitational collapse and space-time singularities},journal={Phys. Rev. Lett.},volume={14},pages={57--59},year={1965}}

@article{hawkingpenrose,author={Hawking, S. W. and Penrose, R.},title={The singularities of gravitational collapse and cosmology},journal={Proc. R. Soc. Lond. A},volume={314},pages={529--548},year={1970}}

@article{israel,author={Israel, Werner},title={Event horizons in static vacuum space-times},journal={Phys. Rev.},volume={164},pages={1776--1779},year={1967}}

@article{carter71,author={Carter, Brandon},title={Axisymmetric black hole has only two degrees of freedom},journal={Phys. Rev. Lett.},volume={26},pages={331--333},year={1971}}

@article{hawking72,author={Hawking, S. W.},title={Black holes in general relativity},journal={Commun. Math. Phys.},volume={25},pages={152--166},year={1972}}

@article{robinson,author={Robinson, D. C.},title={Uniqueness of the {Kerr} black hole},journal={Phys. Rev. Lett.},volume={34},pages={905--906},year={1975}}

@article{carter68,author={Carter, Brandon},title={Global structure of the {Kerr} family of gravitational fields},journal={Phys. Rev.},volume={174},pages={1559--1571},year={1968}}

@article{einstein36,author={Einstein, Albert},title={Lens-like action of a star by the deviation of light in the gravitational field},journal={Science},volume={84},pages={506--507},year={1936}}

@article{darwin,author={Darwin, C.},title={The gravity field of a particle},journal={Proc. R. Soc. Lond. A},volume={249},pages={180--194},year={1959}}

@article{synge,author={Synge, J. L.},title={The escape of photons from gravitationally intense stars},journal={Mon. Not. R. Astron. Soc.},volume={131},pages={463--466},year={1966}}

@incollection{bardeen73,author={Bardeen, James M.},title={Timelike and null geodesics in the {Kerr} metric},booktitle={Black Holes (Les Astres Occlus)},editor={DeWitt, C. and DeWitt, B. S.},publisher={Gordon and Breach},address={New York},pages={215--239},year={1973}}

@article{bpt72,author={Bardeen, James M. and Press, William H. and Teukolsky, Saul A.},title={Rotating black holes: locally nonrotating frames, energy extraction, and scalar synchrotron radiation},journal={Astrophys. J.},volume={178},pages={347--369},year={1972}}

@article{cunninghambardeen,author={Cunningham, C. T. and Bardeen, J. M.},title={The optical appearance of a star orbiting an extreme {Kerr} black hole},journal={Astrophys. J.},volume={183},pages={237--264},year={1973}}

@article{luminet,author={Luminet, J.-P.},title={Image of a spherical black hole with thin accretion disk},journal={Astron. Astrophys.},volume={75},pages={228--235},year={1979}}

@article{ohanian,author={Ohanian, Hans C.},title={The black hole as a gravitational ``lens''},journal={Am. J. Phys.},volume={55},pages={428--432},year={1987}}

@article{falcke00,author={Falcke, Heino and Melia, Fulvio and Agol, Eric},title={Viewing the shadow of the black hole at the {Galactic} center},journal={Astrophys. J. Lett.},volume={528},pages={L13--L16},year={2000}}

@article{teo,author={Teo, Edward},title={Spherical photon orbits around a {Kerr} black hole},journal={Gen. Relativ. Gravit.},volume={35},pages={1909--1926},year={2003}}

@article{devries,author={de Vries, Andreas},title={The apparent shape of a rotating charged black hole, closed photon orbits and the bifurcation set {A4}},journal={Class. Quantum Grav.},volume={17},pages={123--144},year={2000}}

@article{takahashi,author={Takahashi, Rohta},title={Shapes and positions of black hole shadows in accretion disks and spin parameters of black holes},journal={Astrophys. J.},volume={611},pages={996--1004},year={2004}}

@article{beckwithdone,author={Beckwith, Kris and Done, Chris},title={Extreme gravitational lensing near rotating black holes},journal={Mon. Not. R. Astron. Soc.},volume={359},pages={1217--1228},year={2005}}

@article{zakharov05a,author={Zakharov, A. F. and De Paolis, F. and Ingrosso, G. and Nucita, A. A.},title={Direct measurements of black hole charge with future astrometrical missions},journal={Astron. Astrophys.},volume={442},pages={795--799},year={2005}}

@article{zakharov05b,author={Zakharov, A. F. and Nucita, A. A. and De Paolis, F. and Ingrosso, G.},title={Measuring the black hole parameters in the galactic center with {RADIOASTRON}},journal={New Astron.},volume={10},pages={479--489},year={2005}}

@article{bambifreese,author={Bambi, Cosimo and Freese, Katherine},title={Apparent shape of super-spinning black holes},journal={Phys. Rev. D},volume={79},pages={043002},year={2009}}

@article{hiokimaeda,author={Hioki, Kenta and Maeda, Kei-ichi},title={Measurement of the {Kerr} spin parameter by observation of a compact object's shadow},journal={Phys. Rev. D},volume={80},pages={024042},year={2009}}

@article{amarilla10,author={Amarilla, Leonardo and Eiroa, Ernesto F. and Giribet, Gaston},title={Null geodesics and shadow of a rotating black hole in extended {Chern-Simons} modified gravity},journal={Phys. Rev. D},volume={81},pages={124045},year={2010}}

@article{amarilla12,author={Amarilla, Leonardo and Eiroa, Ernesto F.},title={Shadow of a rotating braneworld black hole},journal={Phys. Rev. D},volume={85},pages={064019},year={2012}}

@article{atamurotov13,author={Atamurotov, Farruh and Abdujabbarov, Ahmadjon and Ahmedov, Bobomurat},title={Shadow of rotating non-{Kerr} black hole},journal={Phys. Rev. D},volume={88},pages={064004},year={2013}}

@article{grenzebach,author={Grenzebach, Andreas and Perlick, Volker and L{\"a}mmerzahl, Claus},title={Photon regions and shadows of {Kerr-Newman-NUT} black holes with a cosmological constant},journal={Phys. Rev. D},volume={89},pages={124004},year={2014}}

@article{abdujabbarov16,author={Abdujabbarov, Ahmadjon and Amir, Muhammed and Ahmedov, Bobomurat and Ghosh, Sushant G.},title={Shadow of rotating regular black holes},journal={Phys. Rev. D},volume={93},pages={104004},year={2016}}

@article{tsukamoto18,author={Tsukamoto, Naoki},title={Black hole shadow in an asymptotically flat, stationary, and axisymmetric spacetime: The {Kerr-Newman} and rotating regular black holes},journal={Phys. Rev. D},volume={97},pages={064021},year={2018}}

@article{tsupko17,author={Tsupko, Oleg Yu.},title={Analytical calculation of black hole spin using deformation of the shadow},journal={Phys. Rev. D},volume={95},pages={104058},year={2017}}

@article{perlick15,author={Perlick, Volker and Tsupko, Oleg Yu. and Bisnovatyi-Kogan, Gennady S.},title={Influence of a plasma on the shadow of a spherically symmetric black hole},journal={Phys. Rev. D},volume={92},pages={104031},year={2015}}

@article{perlick17,author={Perlick, Volker and Tsupko, Oleg Yu.},title={Light propagation in a plasma on {Kerr} spacetime: Separation of the {Hamilton-Jacobi} equation and calculation of the shadow},journal={Phys. Rev. D},volume={95},pages={104003},year={2017}}

@article{perlick18,author={Perlick, Volker and Tsupko, Oleg Yu. and Bisnovatyi-Kogan, Gennady S.},title={Black hole shadow in an expanding universe with a cosmological constant},journal={Phys. Rev. D},volume={97},pages={104062},year={2018}}

@article{perlicktsupko22,author={Perlick, Volker and Tsupko, Oleg Yu.},title={Calculating black hole shadows: Review of analytical studies},journal={Phys. Rep.},volume={947},pages={1--39},year={2022}}

@article{cunhaherdeiro18,author={Cunha, Pedro V. P. and Herdeiro, Carlos A. R.},title={Shadows and strong gravitational lensing: a brief review},journal={Gen. Relativ. Gravit.},volume={50},pages={42},year={2018}}

@article{dokuchaev,author={Dokuchaev, Vyacheslav I. and Nazarova, Natalia O.},title={Silhouettes of invisible black holes},journal={Phys. Usp.},volume={63},pages={583--600},year={2020}}

@article{bronzwaerfalcke,author={Bronzwaer, Thomas and Falcke, Heino},title={The nature of black hole shadows},journal={Astrophys. J.},volume={920},pages={155},year={2021}}

@article{narayan19,author={Narayan, Ramesh and Johnson, Michael D. and Gammie, Charles F.},title={The shadow of a spherically accreting black hole},journal={Astrophys. J. Lett.},volume={885},pages={L33},year={2019}}

@article{gralla19,author={Gralla, Samuel E. and Holz, Daniel E. and Wald, Robert M.},title={Black hole shadows, photon rings, and lensing rings},journal={Phys. Rev. D},volume={100},pages={024018},year={2019}}

@article{gralla20a,author={Gralla, Samuel E. and Lupsasca, Alexandru},title={Lensing by {Kerr} black holes},journal={Phys. Rev. D},volume={101},pages={044031},year={2020}}

@article{gralla20b,author={Gralla, Samuel E. and Lupsasca, Alexandru},title={Null geodesics of the {Kerr} exterior},journal={Phys. Rev. D},volume={101},pages={044032},year={2020}}

@article{johnson20,author={Johnson, Michael D. and others},title={Universal interferometric signatures of a black hole's photon ring},journal={Sci. Adv.},volume={6},pages={eaaz1310},year={2020}}

@article{hadar21,author={Hadar, Shahar and Johnson, Michael D. and Lupsasca, Alexandru and Wong, George N.},title={Photon ring autocorrelations},journal={Phys. Rev. D},volume={103},pages={104038},year={2021}}

@article{virbhadraellis00,author={Virbhadra, K. S. and Ellis, G. F. R.},title={{Schwarzschild} black hole lensing},journal={Phys. Rev. D},volume={62},pages={084003},year={2000}}

@article{virbhadraellis02,author={Virbhadra, K. S. and Ellis, G. F. R.},title={Gravitational lensing by naked singularities},journal={Phys. Rev. D},volume={65},pages={103004},year={2002}}

@article{claudel,author={Claudel, Clarissa-Marie and Virbhadra, K. S. and Ellis, G. F. R.},title={The geometry of photon surfaces},journal={J. Math. Phys.},volume={42},pages={818--838},year={2001}}

@article{bozza01,author={Bozza, V. and Capozziello, S. and Iovane, G. and Scarpetta, G.},title={Strong field limit of black hole gravitational lensing},journal={Gen. Relativ. Gravit.},volume={33},pages={1535--1548},year={2001}}

@article{bozza02,author={Bozza, V.},title={Gravitational lensing in the strong field limit},journal={Phys. Rev. D},volume={66},pages={103001},year={2002}}

@article{bozza10,author={Bozza, V.},title={Gravitational lensing by black holes},journal={Gen. Relativ. Gravit.},volume={42},pages={2269--2300},year={2010}}

@article{eiroa02,author={Eiroa, Ernesto F. and Romero, Gustavo E. and Torres, Diego F.},title={{Reissner-Nordstr{\"o}m} black hole lensing},journal={Phys. Rev. D},volume={66},pages={024010},year={2002}}

@article{perlick04,author={Perlick, Volker},title={Gravitational lensing from a spacetime perspective},journal={Living Rev. Relativ.},volume={7},pages={9},year={2004}}

@article{cunha15,author={Cunha, P. V. P. and Herdeiro, C. A. R. and Radu, E. and Runarsson, H. F.},title={Shadows of {Kerr} black holes with scalar hair},journal={Phys. Rev. Lett.},volume={115},pages={211102},year={2015}}

@article{cunha17,author={Cunha, Pedro V. P. and Herdeiro, Carlos A. R. and Radu, Eugen},title={Fundamental photon orbits: black hole shadows and spacetime instabilities},journal={Phys. Rev. D},volume={96},pages={024039},year={2017}}

@article{cunha19,author={Cunha, Pedro V. P. and Herdeiro, Carlos A. R. and Radu, Eugen},title={{EHT} constraint on the ultralight scalar hair of the {M87} supermassive black hole},journal={Universe},volume={5},pages={220},year={2019}}

@article{shaikh18,author={Shaikh, Rajibul and Kocherlakota, Prashant and Narayan, Ramesh and Joshi, Pankaj S.},title={Shadows of spherically symmetric black holes and naked singularities},journal={Mon. Not. R. Astron. Soc.},volume={482},pages={52--64},year={2019}}

@article{shaikhjoshi,author={Shaikh, Rajibul and Joshi, Pankaj S.},title={Can we distinguish black holes from naked singularities by the images of their accretion discs?},journal={J. Cosmol. Astropart. Phys.},volume={2019},number={10},pages={064},year={2019}}

@article{nedkova,author={Nedkova, Petya G. and Tinchev, Vassil K. and Yazadjiev, Stoytcho S.},title={Shadow of a rotating traversable wormhole},journal={Phys. Rev. D},volume={88},pages={124019},year={2013}}

@article{ohgami,author={Ohgami, Takayuki and Sakai, Nobuyuki},title={Wormhole shadows},journal={Phys. Rev. D},volume={91},pages={124020},year={2015}}

@article{abdikamalov,author={Abdikamalov, Askar B. and Abdujabbarov, Ahmadjon A. and Ayzenberg, Dimitry and Malafarina, Daniele and Bambi, Cosimo and Ahmedov, Bobomurat},title={Black hole mimicker hiding in the shadow: Optical properties of the $\gamma$ metric},journal={Phys. Rev. D},volume={100},pages={024014},year={2019}}

@article{kumarghosh20a,author={Kumar, Rahul and Ghosh, Sushant G.},title={Black hole parameter estimation from its shadow},journal={Astrophys. J.},volume={892},pages={78},year={2020}}

@article{kumarghosh20b,author={Kumar, Rahul and Ghosh, Sushant G.},title={Rotating black holes in 4{D} {Einstein-Gauss-Bonnet} gravity and its shadow},journal={J. Cosmol. Astropart. Phys.},volume={2020},number={07},pages={053},year={2020}}

@article{jusufi19,author={Jusufi, Kimet and Jamil, Mubasher and Salucci, Paolo and Zhu, Tao and Haroon, Sohan},title={Black hole surrounded by a dark matter halo in the {M87} galactic center and its identification with shadow images},journal={Phys. Rev. D},volume={100},pages={044012},year={2019}}

@article{konoplya19,author={Konoplya, R. A.},title={Shadow of a black hole surrounded by dark matter},journal={Phys. Lett. B},volume={795},pages={1--6},year={2019}}

@article{khodadi20,author={Khodadi, Mohsen and Allahyari, Alireza and Vagnozzi, Sunny and Mota, David F.},title={Black holes with scalar hair in light of the {Event Horizon Telescope}},journal={J. Cosmol. Astropart. Phys.},volume={2020},number={09},pages={026},year={2020}}

@article{allahyari20,author={Allahyari, Alireza and Khodadi, Mohsen and Vagnozzi, Sunny and Mota, David F.},title={Magnetically charged black holes from non-linear electrodynamics and the {Event Horizon Telescope}},journal={J. Cosmol. Astropart. Phys.},volume={2020},number={02},pages={003},year={2020}}

@article{vagnozzi23,author={Vagnozzi, Sunny and others},title={Horizon-scale tests of gravity theories and fundamental physics from the {Event Horizon Telescope} image of {Sagittarius A*}},journal={Class. Quantum Grav.},volume={40},pages={165007},year={2023}}

@article{bambi19,author={Bambi, Cosimo and Freese, Katherine and Vagnozzi, Sunny and Visinelli, Luca},title={Testing the rotational nature of the supermassive object {M87*} from the circularity and size of its first image},journal={Phys. Rev. D},volume={100},pages={044057},year={2019}}

@article{psaltis20,author={Psaltis, Dimitrios and others},title={Gravitational test beyond the first post-{Newtonian} order with the shadow of the {M87} black hole},journal={Phys. Rev. Lett.},volume={125},pages={141104},year={2020}}

@article{psaltis19,author={Psaltis, Dimitrios},title={Testing general relativity with the {Event Horizon Telescope}},journal={Gen. Relativ. Gravit.},volume={51},pages={137},year={2019}}

@article{johannsenpsaltis10,author={Johannsen, Tim and Psaltis, Dimitrios},title={Testing the no-hair theorem with observations in the electromagnetic spectrum. {I}. Properties of a quasi-{Kerr} spacetime},journal={Astrophys. J.},volume={716},pages={187--197},year={2010}}

@article{johannsen13,author={Johannsen, Tim},title={Photon rings around {Kerr} and {Kerr}-like black holes},journal={Astrophys. J.},volume={777},pages={170},year={2013}}

@article{johannsen16a,author={Johannsen, Tim and others},title={Testing general relativity with the shadow size of {Sgr A*}},journal={Phys. Rev. Lett.},volume={116},pages={031101},year={2016}}

@article{johannsen16b,author={Johannsen, Tim},title={Testing the no-hair theorem with observations of black holes in the electromagnetic spectrum},journal={Class. Quantum Grav.},volume={33},pages={124001},year={2016}}

@article{broderick14,author={Broderick, Avery E. and Johannsen, Tim and Loeb, Abraham and Psaltis, Dimitrios},title={Testing the no-hair theorem with {Event Horizon Telescope} observations of {Sagittarius A*}},journal={Astrophys. J.},volume={784},pages={7},year={2014}}

@article{bambi17,author={Bambi, Cosimo},title={Testing black hole candidates with electromagnetic radiation},journal={Rev. Mod. Phys.},volume={89},pages={025001},year={2017}}

@article{cardosopani,author={Cardoso, Vitor and Pani, Paolo},title={Testing the nature of dark compact objects: a status report},journal={Living Rev. Relativ.},volume={22},pages={4},year={2019}}

@article{cardoso16,author={Cardoso, Vitor and Franzin, Edgardo and Pani, Paolo},title={Is the gravitational-wave ringdown a probe of the event horizon?},journal={Phys. Rev. Lett.},volume={116},pages={171101},year={2016}}

@article{mizuno18,author={Mizuno, Yosuke and others},title={The current ability to test theories of gravity with black hole shadows},journal={Nat. Astron.},volume={2},pages={585--590},year={2018}}

@article{younsi16,author={Younsi, Ziri and Zhidenko, Alexander and Rezzolla, Luciano and Konoplya, Roman and Mizuno, Yosuke},title={New method for shadow calculations: Application to parametrized axisymmetric black holes},journal={Phys. Rev. D},volume={94},pages={084025},year={2016}}

@article{rezzolla14,author={Rezzolla, Luciano and Zhidenko, Alexander},title={New parametrization for spherically symmetric black holes in metric theories of gravity},journal={Phys. Rev. D},volume={90},pages={084009},year={2014}}

@article{konoplya16,author={Konoplya, R. A. and Rezzolla, L. and Zhidenko, A.},title={General parametrization of axisymmetric black holes in metric theories of gravity},journal={Phys. Rev. D},volume={93},pages={064015},year={2016}}

@article{held19,author={Held, Alexander and Gold, Roman and Eichhorn, Astrid},title={Asymptotic safety casts its shadow},journal={J. Cosmol. Astropart. Phys.},volume={2019},number={06},pages={029},year={2019}}

@article{wei19,author={Wei, Shao-Wen and Liu, Yu-Xiao and Mann, Robert B.},title={Intrinsic curvature and topology of shadows in {Kerr} spacetime},journal={Phys. Rev. D},volume={99},pages={041303},year={2019}}

@article{tsupko20,author={Tsupko, Oleg Yu. and Fan, Zuhui and Bisnovatyi-Kogan, Gennady S.},title={Black hole shadow as a standard ruler in cosmology},journal={Class. Quantum Grav.},volume={37},pages={065016},year={2020}}

@article{gyoto,author={Vincent, F. H. and Paumard, T. and Gourgoulhon, E. and Perrin, G.},title={{GYOTO}: a new general relativistic ray-tracing code},journal={Class. Quantum Grav.},volume={28},pages={225011},year={2011}}

@article{harm,author={Gammie, Charles F. and McKinney, Jonathan C. and T{\'o}th, G{\'a}bor},title={{HARM}: A numerical scheme for general relativistic magnetohydrodynamics},journal={Astrophys. J.},volume={589},pages={444--457},year={2003}}

@article{porth19,author={Porth, Oliver and others},title={The {Event Horizon} General Relativistic Magnetohydrodynamic Code Comparison Project},journal={Astrophys. J. Suppl. Ser.},volume={243},pages={26},year={2019}}

@article{gold20,author={Gold, Roman and others},title={Verification of radiative transfer schemes for the {EHT}},journal={Astrophys. J.},volume={897},pages={148},year={2020}}

@article{dexter10,author={Dexter, Jason and Agol, Eric and Fragile, P. Chris and McKinney, Jonathan C.},title={The submillimeter bump in {Sgr A*} from relativistic {MHD} simulations},journal={Astrophys. J.},volume={717},pages={1092--1104},year={2010}}

@article{moscibrodzka16,author={Mo{\'s}cibrodzka, Monika and Falcke, Heino and Shiokawa, Hotaka},title={General relativistic magnetohydrodynamical simulations of the jet in {M87}},journal={Astron. Astrophys.},volume={586},pages={A38},year={2016}}

@article{shakura,author={Shakura, N. I. and Sunyaev, R. A.},title={Black holes in binary systems. Observational appearance},journal={Astron. Astrophys.},volume={24},pages={337--355},year={1973}}

@article{narayanyi,author={Narayan, Ramesh and Yi, Insu},title={Advection-dominated accretion: a self-similar solution},journal={Astrophys. J. Lett.},volume={428},pages={L13--L16},year={1994}}

@article{yuannarayan,author={Yuan, Feng and Narayan, Ramesh},title={Hot accretion flows around black holes},journal={Annu. Rev. Astron. Astrophys.},volume={52},pages={529--588},year={2014}}

@article{abramowicz,author={Abramowicz, Marek A. and Fragile, P. Chris},title={Foundations of black hole accretion disk theory},journal={Living Rev. Relativ.},volume={16},pages={1},year={2013}}

@article{bz77,author={Blandford, R. D. and Znajek, R. L.},title={Electromagnetic extraction of energy from {Kerr} black holes},journal={Mon. Not. R. Astron. Soc.},volume={179},pages={433--456},year={1977}}

@article{blandford19,author={Blandford, Roger and Meier, David and Readhead, Anthony},title={Relativistic jets from active galactic nuclei},journal={Annu. Rev. Astron. Astrophys.},volume={57},pages={467--509},year={2019}}

@article{broderickloeb05,author={Broderick, Avery and Loeb, Abraham},title={Imaging bright spots in the accretion flow near the black hole horizon of {Sgr A*}},journal={Mon. Not. R. Astron. Soc.},volume={363},pages={353--362},year={2005}}

@article{broderickloeb06,author={Broderick, Avery E. and Loeb, Abraham},title={Imaging optically thin hot spots near the black hole horizon of {Sgr A*} at radio and near-infrared wavelengths},journal={Mon. Not. R. Astron. Soc.},volume={367},pages={905--916},year={2006}}

@article{doeleman08,author={Doeleman, Sheperd S. and others},title={Event-horizon-scale structure in the supermassive black hole candidate at the {Galactic} Centre},journal={Nature},volume={455},pages={78--80},year={2008}}

@article{doeleman12,author={Doeleman, Sheperd S. and others},title={Jet-launching structure resolved near the supermassive black hole in {M87}},journal={Science},volume={338},pages={355--358},year={2012}}

@article{akiyama15,author={Akiyama, Kazunori and others},title={230 {GHz} {VLBI} observations of {M87}: event-horizon-scale structure during an enhanced very-high-energy $\gamma$-ray state in 2012},journal={Astrophys. J.},volume={807},pages={150},year={2015}}

@article{akiyama17,author={Akiyama, Kazunori and others},title={Imaging the {Schwarzschild}-radius-scale structure of {M87} with the {Event Horizon Telescope} using sparse modeling},journal={Astrophys. J.},volume={838},pages={1},year={2017}}

@article{chael18,author={Chael, Andrew A. and others},title={Interferometric imaging directly with closure phases and closure amplitudes},journal={Astrophys. J.},volume={857},pages={23},year={2018}}

@article{lu23,author={Lu, Ru-Sen and others},title={A ring-like accretion structure in {M87} connecting its black hole and jet},journal={Nature},volume={616},pages={686--690},year={2023}}

@article{ghez08,author={Ghez, A. M. and others},title={Measuring distance and properties of the {Milky Way}'s central supermassive black hole with stellar orbits},journal={Astrophys. J.},volume={689},pages={1044--1062},year={2008}}

@article{genzel10,author={Genzel, Reinhard and Eisenhauer, Frank and Gillessen, Stefan},title={The Galactic Center massive black hole and nuclear star cluster},journal={Rev. Mod. Phys.},volume={82},pages={3121--3195},year={2010}}

@article{gravity18,author={{GRAVITY Collaboration}},title={Detection of the gravitational redshift in the orbit of the star {S2} near the {Galactic} centre massive black hole},journal={Astron. Astrophys.},volume={615},pages={L15},year={2018}}

@article{gw150914,author={Abbott, B. P. and others},title={Observation of gravitational waves from a binary black hole merger},journal={Phys. Rev. Lett.},volume={116},pages={061102},year={2016}}

@article{gebhardt11,author={Gebhardt, Karl and Adams, Joshua and Richstone, Douglas and others},title={The black hole mass in {M87} from {Gemini}/{NIFS} adaptive optics observations},journal={Astrophys. J.},volume={729},pages={119},year={2011}}

@article{walsh13,author={Walsh, Jonelle L. and Barth, Aaron J. and Ho, Luis C. and Sarzi, Marc},title={The {M87} black hole mass from gas-dynamical models of {Space Telescope Imaging Spectrograph} observations},journal={Astrophys. J.},volume={770},pages={86},year={2013}}

@article{johnson23,author={Johnson, Michael D. and others},title={Key science goals for the next-generation {Event Horizon Telescope}},journal={Galaxies},volume={11},pages={61},year={2023}}

@article{doeleman23,author={Doeleman, Sheperd S. and others},title={Reference array and design considerations for the next-generation {Event Horizon Telescope}},journal={Galaxies},volume={11},pages={107},year={2023}}

@article{roelofs19,author={Roelofs, Freek and others},title={Simulations of imaging the event horizon of {Sagittarius A*} from space},journal={Astron. Astrophys.},volume={625},pages={A124},year={2019}}

@article{wielgus22,author={Wielgus, Maciek and others},title={Millimeter light curves of {Sagittarius A*} observed during the 2017 {Event Horizon Telescope} campaign},journal={Astrophys. J. Lett.},volume={930},pages={L19},year={2022}}

@book{chandra,author={Chandrasekhar, S.},title={The Mathematical Theory of Black Holes},publisher={Oxford University Press},address={Oxford},year={1983}}

@book{mtw,author={Misner, Charles W. and Thorne, Kip S. and Wheeler, John A.},title={Gravitation},publisher={W. H. Freeman},address={San Francisco},year={1973}}

@book{wald,author={Wald, Robert M.},title={General Relativity},publisher={University of Chicago Press},address={Chicago},year={1984}}

@book{frolovnovikov,author={Frolov, Valeri P. and Novikov, Igor D.},title={Black Hole Physics: Basic Concepts and New Developments},publisher={Kluwer Academic},address={Dordrecht},year={1998}}

@article{vagnozzi20ruler,author={Vagnozzi, Sunny and Bambi, Cosimo and Visinelli, Luca},title={Concerns regarding the use of black hole shadows as standard rulers},journal={Class. Quantum Grav.},volume={37},pages={087001},year={2020},eprint={2001.02986},archivePrefix={arXiv}}

@article{li14,author={Li, Zilong and Bambi, Cosimo},title={Measuring the {Kerr} spin parameter of regular black holes from their shadow},journal={J. Cosmol. Astropart. Phys.},volume={2014},number={01},pages={041},year={2014},eprint={1309.1606},archivePrefix={arXiv}}

@article{ghasemi15,author={Ghasemi-Nodehi, M. and Li, Zilong and Bambi, Cosimo},title={Shadows of {CPR} black holes and tests of the {Kerr} metric},journal={Eur. Phys. J. C},volume={75},pages={315},year={2015},eprint={1506.02627},archivePrefix={arXiv}}

@article{gralla21,author={Gralla, Samuel E.},title={Can the {EHT} {M87} results be used to test general relativity?},journal={Phys. Rev. D},volume={103},pages={024023},year={2021},eprint={2010.08557},archivePrefix={arXiv}}

@article{gralla20c,author={Gralla, Samuel E. and Lupsasca, Alexandru and Marrone, Daniel P.},title={The shape of the black hole photon ring: A precise test of strong-field general relativity},journal={Phys. Rev. D},volume={102},pages={124004},year={2020},eprint={2008.03879},archivePrefix={arXiv}}

@article{himwich20,author={Himwich, Elizabeth and Johnson, Michael D. and Lupsasca, Alexandru and Strominger, Andrew},title={Universal polarimetric signatures of the black hole photon ring},journal={Phys. Rev. D},volume={101},pages={084020},year={2020},eprint={2001.08750},archivePrefix={arXiv}}

@article{kocherlakota22,author={Kocherlakota, Prashant and Rezzolla, Luciano},title={Distinguishing gravitational and emission physics in black-hole imaging: spherical symmetry},journal={Mon. Not. R. Astron. Soc.},volume={513},pages={1229},year={2022},eprint={2201.05641},archivePrefix={arXiv}}

@article{stella99,author={Stella, Luigi and Vietri, Mario},title={k{H}z quasi-periodic oscillations in low-mass {X}-ray binaries as probes of general relativity in the strong-field regime},journal={Phys. Rev. Lett.},volume={82},pages={17--20},year={1999},eprint={astro-ph/9812124},archivePrefix={arXiv}}

@article{motta14,author={Motta, S. E. and Belloni, T. M. and Stella, L. and Mu{\~n}oz-Darias, T. and Fender, R.},title={Precise mass and spin measurements for a stellar-mass black hole through {X}-ray timing: the case of {GRO J1655-40}},journal={Mon. Not. R. Astron. Soc.},volume={437},pages={2554--2565},year={2014},eprint={1309.3652},archivePrefix={arXiv}}

@article{bueno16,author={Bueno, Pablo and Cano, Pablo A.},title={Four-dimensional black holes in {Einsteinian} cubic gravity},journal={Phys. Rev. D},volume={94},pages={124051},year={2016},eprint={1610.08019},archivePrefix={arXiv}}

@article{cano20,author={Cano, Pablo A. and Fransen, Kwinten and Hertog, Thomas},title={Ringing of rotating black holes in higher-derivative gravity},journal={Phys. Rev. D},volume={102},pages={044047},year={2020},eprint={2005.03671},archivePrefix={arXiv}}

@article{cano22,author={Cano, Pablo A. and Fransen, Kwinten and Hertog, Thomas and Maenaut, Simon},title={Gravitational ringing of rotating black holes in higher-derivative gravity},journal={Phys. Rev. D},volume={105},pages={024064},year={2022},eprint={2110.11378},archivePrefix={arXiv}}

@article{endlich17,author={Endlich, Solomon and Gorbenko, Victor and Huang, Junwu and Senatore, Leonardo},title={An effective formalism for testing extensions to general relativity with gravitational waves},journal={J. High Energy Phys.},volume={2017},number={09},pages={122},year={2017},eprint={1704.01590},archivePrefix={arXiv}}

@article{derham20,author={de Rham, Claudia and Francfort, J{\'e}r{\'e}mie and Zhang, Jun},title={Black hole gravitational waves in the effective field theory of gravity},journal={Phys. Rev. D},volume={102},pages={024079},year={2020}}

@article{berti15,author={Berti, Emanuele and others},title={Testing general relativity with present and future astrophysical observations},journal={Class. Quantum Grav.},volume={32},pages={243001},year={2015},eprint={1501.07274},archivePrefix={arXiv}}

@article{cardosogualtieri16,author={Cardoso, Vitor and Gualtieri, Leonardo},title={Testing the black hole `no-hair' hypothesis},journal={Class. Quantum Grav.},volume={33},pages={174001},year={2016},eprint={1607.03133},archivePrefix={arXiv}}

@article{burgess04,author={Burgess, C. P.},title={Quantum gravity in everyday life: General relativity as an effective field theory},journal={Living Rev. Relativ.},volume={7},pages={5},year={2004},eprint={gr-qc/0311082},archivePrefix={arXiv}}

@article{tsukamoto14,
	author = "Tsukamoto, Naoki and Li, Zilong and Bambi, Cosimo",
	title = "{Constraining the spin and the deformation parameters from the black hole shadow}",
	eprint = "1403.0371",
	archivePrefix = "arXiv",
	primaryClass = "gr-qc",
	doi = "10.1088/1475-7516/2014/06/043",
	journal = "JCAP",
	volume = "06",
	pages = "043",
	year = "2014"
}

@article{Hennigar17,
	author = "Hennigar, Robie A. and Mann, Robert B.",
	title = "{Black holes in Einsteinian cubic gravity}",
	eprint = "1610.06675",
	archivePrefix = "arXiv",
	primaryClass = "hep-th",
	doi = "10.1103/PhysRevD.95.064055",
	journal = "Phys. Rev. D",
	volume = "95",
	number = "6",
	pages = "064055",
	year = "2017"
}

@misc{chael18heating,author={Chael, Andrew and Rowan, Michael and Narayan, Ramesh and Johnson, Michael and Sironi, Lorenzo},title={The role of electron heating physics in images and variability of the {Galactic Centre} black hole {Sagittarius A*}},year={2018},eprint={1804.06416},archivePrefix={arXiv},note={Journal reference to be confirmed}}

@article{Wei13,
	author = "Wei, Shao-Wen and Liu, Yu-Xiao",
	title = "{Observing the shadow of Einstein-Maxwell-Dilaton-Axion black hole}",
	eprint = "1311.4251",
	archivePrefix = "arXiv",
	primaryClass = "gr-qc",
	doi = "10.1088/1475-7516/2013/11/063",
	journal = "JCAP",
	volume = "11",
	pages = "063",
	year = "2013"
}

@article{lee23,author={Lee, Tsai-Chen and Hu, Zezhou and Guo, Minyong and Chen, Bin},title={Circular orbits and polarized images of charged particles orbiting {Kerr} black hole with a weak magnetic field},journal={Phys. Rev. D},volume={108},pages={024008},year={2023},eprint={2211.04143},archivePrefix={arXiv}}

@article{Peng,
	author = "Peng, Jun and Guo, Minyong and Feng, Xing-Hui",
	title = "{Influence of quantum correction on black hole shadows, photon rings, and lensing rings}",
	eprint = "2008.00657",
	archivePrefix = "arXiv",
	primaryClass = "gr-qc",
	doi = "10.1088/1674-1137/ac06bb",
	journal = "Chin. Phys. C",
	volume = "45",
	number = "8",
	pages = "085103",
	year = "2021"
}

@article{Meng:2024puu,
	author = "Meng, Yuan and Kuang, Xiao-Mei and Wang, Xi-Jing and Wang, Bin and Wu, Jian-Pin",
	title = {{Images of hairy Reissner{\textendash}Nordstr{\"o}m black hole illuminated by static accretions}},
	eprint = "2401.05634",
	archivePrefix = "arXiv",
	primaryClass = "gr-qc",
	doi = "10.1140/epjc/s10052-024-12686-w",
	journal = "Eur. Phys. J. C",
	volume = "84",
	number = "3",
	pages = "305",
	year = "2024"
}

@article{Qin:2022kaf,
	author = "Qin, Xin and Chen, Songbai and Zhang, Zelin and Jing, Jiliang",
	title = "{Polarized Image of a Rotating Black Hole in Scalar{\textendash}Tensor{\textendash}Vector{\textendash}Gravity Theory}",
	eprint = "2207.12034",
	archivePrefix = "arXiv",
	primaryClass = "gr-qc",
	doi = "10.3847/1538-4357/ac8f49",
	journal = "Astrophys. J.",
	volume = "938",
	number = "1",
	pages = "2",
	year = "2022"
}

@article{Chen:2024cxi,
	author = "Chen, Yiqian and Cheng, Lang and Wang, Peng and Yang, Haitang",
	title = "{Polarized image of a synchrotron-emitting ring in an Einstein-Maxwell-scalar theory}",
	eprint = "2409.05304",
	archivePrefix = "arXiv",
	primaryClass = "gr-qc",
	doi = "10.1103/kbkh-xgp7",
	journal = "Phys. Rev. D",
	volume = "112",
	number = "2",
	pages = "024044",
	year = "2025"
}

@article{Cano:2019ore,
    author = "Cano, Pablo A. and Ruip{\'e}rez, Alejandro",
    title = "{Leading higher-derivative corrections to Kerr geometry}",
    eprint = "1901.01315",
    archivePrefix = "arXiv",
    primaryClass = "gr-qc",
    reportNumber = "IFT-UAM/CSIC-19-2",
    doi = "10.1007/JHEP05(2019)189",
    journal = "JHEP",
    volume = "05",
    pages = "189",
    year = "2019",
    note = "[Erratum: JHEP 03, 187 (2020)]"
}

\end{document}